\documentclass[a4paper,fleqn,usenatbib]{mnras}

\usepackage{graphicx}	
\usepackage{amsmath}	
\usepackage{amssymb}	
\usepackage{mwe}
\usepackage{multicol}

\title[Hydrodynamic simulations of SECCO~1]{Hydrodynamic simulations of an isolated star-forming gas cloud in the Virgo cluster}
\author[Calura et al.]
{Francesco Calura$^{1}$\thanks{E-mail: francesco.calura@inaf.it}, Michele  Bellazzini$^{1}$, Annibale D'Ercole$^{1}$
\\ ~ \\
$^{1}$INAF-Osservatorio di Astrofisica e Scienza dello Spazio di Bologna, Via Gobetti 93/3, I-40129 Bologna, Italy
}

\begin{document}
\label{firstpage}
\pagerange{\pageref{firstpage}--\pageref{lastpage}} \pubyear{2019}

\maketitle

\begin{abstract}
We present a suite of three-dimensional, high-resolution hydrodynamic
simulations that follow the  evolution of a massive ($10^7~M_{\sun}$)
pressure confined, star-forming neutral gas cloud moving through a hot
intra-cluster medium (ICM).  The main goal of the analysis is to get
theoretical insight into the lifetimes and evolution of stellar
systems like the recently discovered star-forming cloud SECCO~1 in the
Virgo cluster of galaxies, but it may be of general interest for the
study of the star forming gas clumps that are observed in the tails of
ram pressure stripped galaxies. 
Building up on a previous, simple simulation, we explored the effect
of different relative velocity of the cloud and larger temperature of
the ICM,  as well as the effect of the cloud self-gravity. Moreover,
we  performed a simulation including star-formation and stellar
feedback, allowing for a first time a direct comparison with the
observed properties of the stars in the system.  The survivability of
the cold gas in the simulated clouds is granted on timescales of the
order of 1 Gyr, with final cold gas fractions generally $>0.75$.  In
all cases,  the simulated systems end up, after 1 Gyr of evolution, as
symmetric clouds in pressure equilibrium with the external hot gas.
We also confirm that gravity played a negligible role at the largest
scales on the evolution of the clouds.  In our simulation with star
formation, star formation begins immediately, it peaks at 
the earliest times and decreases monotonically with time.
Inhomogeneous supernova explosions are the cause of an asymmetric
shape of the gas cloud, facilitating the development of instabilities
and the decrease of the cold gas fraction. 
\end{abstract}

\begin{keywords}
Galaxies: star formation - methods: numerical - hydrodynamics - galaxies: individual. 
\end{keywords}



\section{Introduction} 
Within a survey aimed at looking for stellar counterparts in compact
HI clouds\footnote{SECCO survey, see {\tt  http://www.bo.astro.it/secco}},
\cite{pap1} found ongoing star formation within an isolated gas cloud lying in the Virgo cluster (see also \citealt{secco1}). 
The mass budget of this stellar system, named SECCO~1, is dominated by neutral hydrogen 
($M_{\rm HI}\sim10^7 M_\odot$), while the stellar mass is $M_*\sim10^5 M_\odot$, (see also \citealt{sand15}, 2017, and \citealt{adams15}). 
The nebular metallicity as measured in several HII regions within the system is  12+log(O/H)=8.38$\pm$ 0.11,  much higher than that of
dwarf galaxies of similar mass, implying that the cloud could have been stripped from a larger galaxy
(\citealt{pap_muse}, \citealt{bel18}, B18 hereafter), but, in fact, SECCO~1 lies several hundreds of kpc apart from any known 
galaxy within the Virgo cluster (see \citealt{sand17} and B18). 

Indeed, all the known properties of SECCO~1 lie in the range covered by star-forming gas clumps observed in the extended tails of ram pressure stripped gas in the
so-called {\em jellyfish} galaxies (\citealt{pog19} and references therein). The crucial difference with these systems is the extreme isolation of SECCO~1.
This cloud could have travelled away from its parent galaxy for a long time ($\ga 1$~Gyr) within the Intra Cluster Medium (ICM) of Virgo, preserving  the conditions to 
form stars today, in spite of not being gravitationally bound (B18). \cite{sand17} presented a list of five additional candidate siblings 
of SECCO~1. Hence, SECCO~1 may  be the prototype of a new class of stellar systems. 

By means of hydrodynamical simulations, \cite{kap09} showed that in galaxy clusters, 
ram-pressure stripping episodes 
are expected to extract from disc galaxies a large number of  star-forming blobs,
and that such blobs are expected to be long-lived in the cluster medium. 
In the past, other models have been proposed 
(\citealt{bek05}) which showed that large amounts of neutral gas could be removed from the parent galaxy by tidal encounters. 
\cite{bur16} suggested that small clumps stripped from gas-rich
systems can travel for a relatively long time, confined by the pressure of the surrounding hot ICM. They showed that pressure confinement may drive them to density conditions conducive to the ignition of star formation. 

In an attempt to verify the plausibility of the survival over long travel times of a cloud like SECCO~1,  B18  performed both  two-dimensional (2D) and 
three-dimensional (3D) hydrodynamical simulations of an initially cold ($T\sim5000$ K) gas cloud moving subsonically (but with a large speed, $\sim$200 km/s) 
through a very hot ($T\sim$ a few 10$^6$ K) medium.  
Their results indicated that the cloud was able to preserve $\simeq$75\% of its initial content of cold gas during its 1~Gyr -long flight, 
while the remaining $\simeq$25\% was evaporated into the hot medium. 
The cloud also experienced a considerable morphological evolution, mostly driven by Kelvin-Helmholtz instabilities (KHI). 

However, important physical ingredients were missing in the B18 simulations, including, e.g., the self-gravity of the cloud. 
Additional numerical experiments are needed to 
test further the role of the environment in determining the survivability of the cloud. 
In particular, how does the thermal state of the ICM affect the evolution of the cloud? What is the effect of a different relative
velocity between SECCO~1 and the ICM in its long-term evolution? 
Last but not least, what are the effects of star formation and stellar feedback? 
In order to address these issues, in this paper we present a new suite of high-resolution, 3D simulations aimed at describing the motion of a cold gas cloud similar to SECCO~1 through a hot ICM. \\
A description of the code, of the initial setup and of the physical ingredients of our simulations are presented in Sec.~\ref{sec_model}.
In in Sec.~\ref{sec_results}, our results are presented and discussed.
Finally, we draw our conclusions in Sect. ~\ref{sec_conclusions}.

\section{Model description} \label{sec_model}
\begin{figure*}
\includegraphics[width=14.5cm]{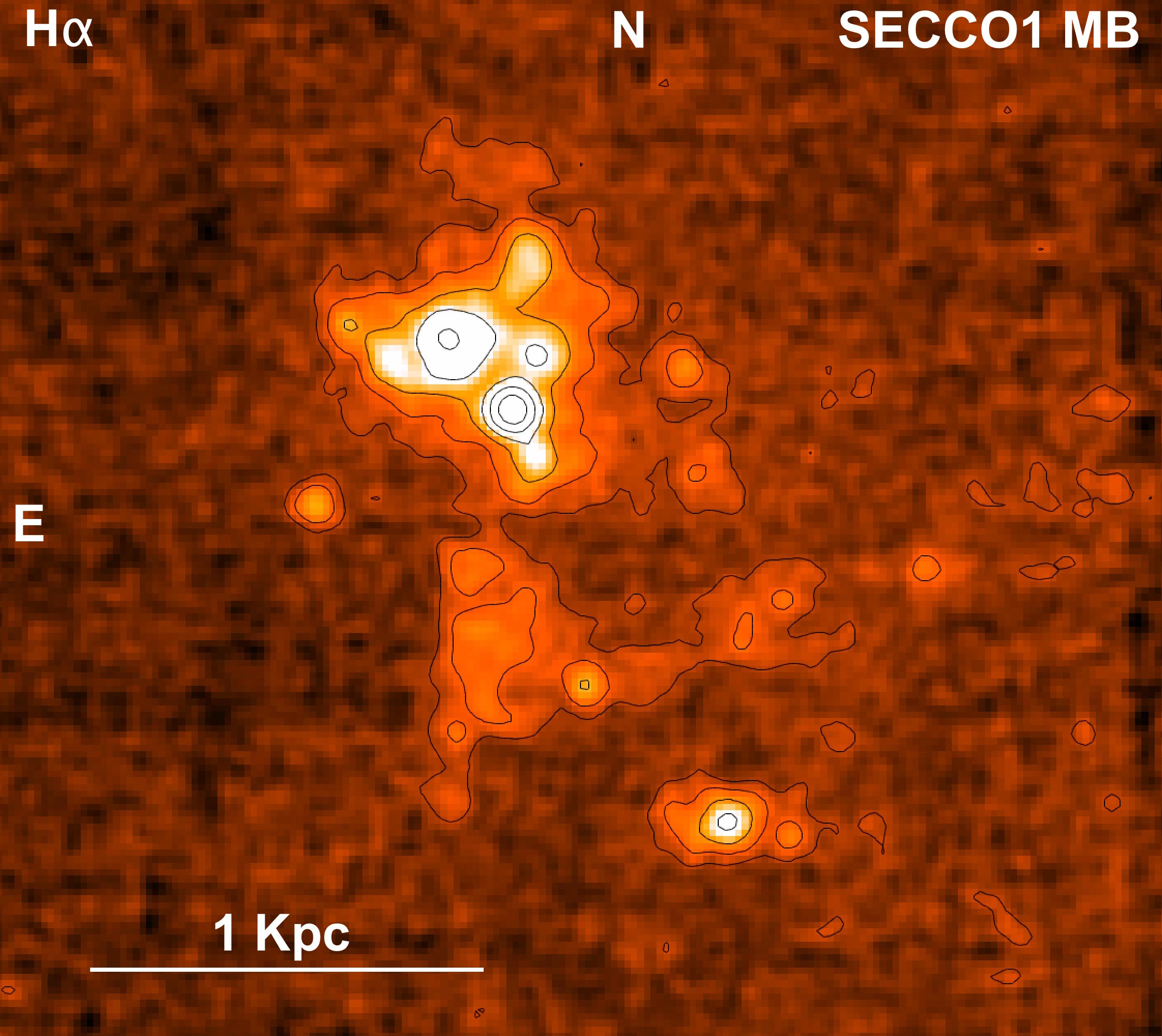}
\caption{H$\alpha$ image of the main body of SECCO 1 as obtained from MUSE data (see \citealt{pap_muse}). 
The black contour lines represent the flux  at  %
0.5, 1.0, 2.0, 4.0,  8.0 and 16.0 $\times 10^{-18}$ erg cm$^{-2}$ s$^{-1}$.}
\label{fig_secco}
\end{figure*}

Our numerical simulations are designed to describe the motion of a cold gas cloud
with respect to a background distribution of hot gas, representing the Virgo ICM. 
The cold gas cloud is assumed to be initially spherical, and it is designed to represent the main body of SECCO~1,
which contains a large fraction ($\sim 80 \%$) of the total gas mass detected in the entire system (B18), 
which also includes a secondary body, which accounts for the remainder $\sim20\%$ of the mass.
An H$\alpha$ intensity map of the main body of SECCO 1 is presented in Fig.~\ref{fig_secco}, whereas a summary of its main 
observational properies is in Tab.~\ref{tab_obs}.

In the present work, we do not investigate how the 
cold gas cloud can have originated. 
We study its evolution after it has abandoned its parent galaxy
and is outside the effects of its gravitational field.
We focus on a parameter space which includes a few fundamental
properties of system and of the external medium, including temperature and relative velocity. 
After some pure hydrodynamic, radiative runs, we also perform runs which take into account
the self-gravity of the gas. 
Finally, in the most refined model, we consider star formation 
and stellar feedback and study their effects on the evolution of the gas cloud. 

It is important to stress that 
our idealised, numerical experiments are designed 
to gain an insight into the physical processes that are expected to play a major role in the evolution of this kind
of systems. 

\begin{table*}
\caption{Summary of the main observational properties of the main body of SECCO.}
\begin{tabular}{ccc}
\hline
Quantity            & SECCO Main Cloud   & References  \\
                    & Observed value or range     &             \\  
\hline
Distance [Mpc]             &    17                   & \cite{bos14}         \\
$M_{*}$ [$M_{\odot}$]       &   4 $\times 10^4$ -  1.2 $\times 10^5$$^a$                &   \cite{sand17}, \cite{pap_muse}          \\
$M_{H~I}$ [$M_{\odot}$]     &   1.5 $\times 10^7$                 &     \cite{adams15}        \\
$R_{*}$ [kpc]              &   1.2                   &       \cite{pap_muse}      \\
$R_{\rm HI}$ [kpc]             &   3.7                  &        \cite{adams15}     \\
$\sigma_{*}$ [km~s$^{-1}$]  &    3.5$\pm2.1$               &     \cite{pap_muse}         \\
$\sigma_{\rm HI}$ [km~s$^{-1}$]  &   9 $\pm3$               &     \cite{adams15}        \\
SFR [$M_{\odot}$~yr$^{-1}$]  &        $\sim$1 $\times 10^{-3}$ $^b$          &   \cite{sand17}         \\
\hline
\hline
\end{tabular}
\begin{tabular}{l}
$^{a}$ The reported range includes the value derived from  CMD analysis and the upper limit derived from the integrated light. \\
$^{b}$ Derived from CMD analysis.\\
\end{tabular}
\label{tab_obs}
\end{table*}

\subsection{Simulation Setup} 
The code used is a customized version of the grid-based, Adaptive Mesh Refinement (AMR) hydro-code RAMSES (\citealt{tey02}).
The AMR technique allows one to reduce the computational cost of a simulation by
assigning higher resolution to zones of particular interest within the computational box.
In our case, the AMR feature is exploited by adopting both geometry-based and discontinuity-based
criteria. 
In our Cartesian Grid, the computational box has a volume of $L_{box}^{3}=(40$ kpc$)^3$,
chosen in order to largely contain the whole cloud and, as much as possible, the extent of instabilities.
For a given setup, in general more simulations are run at various resolution
and starting from the same initial conditions, in order to check numerical convergence. 
The maximum spatial resolution used in our simulations is of 76 pc.\\
As in B18, the initial radius of the spherical gas cloud is $\sim 3.7$ kpc.
With this configuration, the side of the box is $\sim 10$ times the initial radius of the cloud. \\
Our setup is designed to recreate the fast motion of the cloud in the hot ICM of the Virgo cluster. 
In our 'wind tunnel' numerical experiment (with a basic setup similar to the one of \citealt{cal19}), 
the reference frame is placed at the centre of the cold gas cloud, and since the start of the
simulation the hot ICM gas 
moves continuously along the x-axis.  
New fast, hot gas is continuously allowed to flow inside our computational box from one of the boundaries and again
moving parallel to the x-axis, from left to right.\\
The parameters describing the initial conditions of the models investigated in this paper are reported in Table~\ref{tab_models} (columns 2-8).

\begin{table*}
\flushleft
  \caption{Parameters defining the initial conditions for the models investigated in our simulations.
  First colums is model name, second column is cloud density, third column is cloud temperature,
  fourth column is density of the ICM, fifth column is temperature of the ICM,
  sixth column is the cloud velocity. Seventh and eighth columns indicate whether gravity and star formation are activated in the
  model, respectively. In the last three columns we report a few quantities estimated
  for each model, namely the cloud half-mass radius (nineth column), line width (tenth column) and SFR (eleventh column).}
 \begin{tabular}{cccccccc|ccc}
\hline
Model     & &  &   Adopted value           &         &   &  &    & & Computed quantities &  \\
          & &  &                           &         &   &  &    & & (0.25 Gyr -1 Gyr) &  \\
\hline
\tiny          & $\rho_{cl}$         & $T_{cl}$ & $\rho_{ICM}$            & $T_{ICM}$        & $v_{cl}$  & Gravity & SF &$R_{c, hm}$  & $\sigma_l$ & $SFR$ \\
          & { $(g~cm^{ -3})$}          &  {$(K)$}  & { $(g~cm^{-3})$}  & { $(K)$}           & { (km~s$^{-1}$)} &   &       &  { (kpc)}  &  { (km~s$^{-1}$)} &  { (10$^{-3}$$M_{\odot}$yr$^{-1}$)}       \\
\hline
Standard   &  3$\times 10^{-26}$ &  $10^4$  & 2.5$\times10^{-29}$   & $5\times10^6$ &  200      &  N      & N    &3.3-6.4& $<20.8$  & -\\
T1E7K      &  3$\times 10^{-26}$ &  $10^4$  & 2.5$\times10^{-29}$   & $10^7$          &  200      &  N      & N   &3.1-4.2& $<19.5$  &-\\
Supersonic   &  3$\times 10^{-26}$ &  $10^4$  & 2.5$\times10^{-29}$   & $5\times10^6$ &  400      &  N      & N  &3.1-4.4& $<22.6$  &-\\
Gravity     &  3$\times 10^{-26}$ &  $10^4$  & 2.5$\times10^{-29}$   & $5\times10^6$ &  200      &  Y      & N   &2.7-5.8& $<21.8$  &-\\
\hline
Star Formation     &  3$\times 10^{-26}$ &  $10^4$  & 2.5$\times10^{-29}$   & $5\times10^6$ &  200      &  Y      & Y   &3.5-6.2& $<20.2$ &0.6-2.5\\
Supersonic+      &  3$\times 10^{-26}$ &  $10^4$  & 2.5$\times10^{-29}$   & $5\times10^6$ &  400      &  Y      & Y  &3.1-4.2 & $<23.6$  &0.6-2.5\\
Star Formation   &                    &          &                       &                 &           &         &    &        &          &  \\
\hline
\hline	      
\end{tabular} 
\label{tab_models}
\end{table*}

The parameters of our Standard model are the same as the ones of the simulations 
carried on in B18, with an initial density of the cloud and of the ICM of 3$\times 10^{-26}$   and 2.5$\times~10^{-29}~g/cm^3$ respectively.
As discussed in B18, the parameters of the Standard run were derived from the available observations. 
The temperature of the hot gas $T_{ICM}$ was derived from the radial velocity dispersion of the ICM as determined by \cite{bos14}.  
The density of the ICM was derived from the total mass (dark and baryonic matter) of the
low velocity cloud  substructure of the Virgo cluster, assuming that the hot gas density distribution
follows the dark matter density profile \citep{nav96}. 
As for the cloud, for its initial temperature a value comparable to  the mean temperature of the diffuse HI was assumed
(e.g. \citealt{wol03}), whereas its density was computed from the observed radius and mass of SECCO~1 reported in Tab.~\ref{tab_obs}
and considering a mass abundance of HI with respect to all elements at the temperature of the cloud of $\sim 0.16$ (B18). 
Note that, with this choice, at the beginning of the simulation the cloud is almost 
in pressure equilibrium with the surrounding environment.

It may be interesting to estimate how far out of equilibrium the cloud is initially,
e. g. by computing the radius it needs to have to be in equilibrium. 
This has important consequences for the suggestion of \cite{bur16} that pressure confinement
can explain the survival of clouds similar to SECCO~1.  
This was tested also in \cite{tay18}, but only in clouds where dynamic pressure
dominates over thermal pressure. 
The much lower line width of the cloud explored here potentially gives this study greater similarity
to the original pressure confinement model. 
The equilibrium radius can be estimated from the initial conditions shown in Tab.~\ref{tab_models},
suggesting that the ratio between the intial pressure
of the cloud of the stardard run and the one of the hot ICM is of 2.4. 
Considering $\rho_{cl} \propto R_{\rm HI}^{-3}$, a condition of complete equilibrium requires
a radius of the cloud $\sim 1.3 R_{\rm HI}$. 
The model in which the deviation from initial pressure equilibrium is the smallest
is the T1E7K,
with an initial ratio between cloud pressure and ICM pressure of 1.2. 
   
The default value chosen for the relative velocity between the cloud and the ICM is  200 km/s, but we will also test a
larger value (400 km/s). 
Initially, both the cold cloud and the external medium are homogeneous and characterised by a uniform temperature.
The cloud and the ICM have uniform initial metallicity values of $0.5 Z_{\odot}$ and $0.1 Z_{\odot}$, respectively.
Initially, the isothermal cloud temperature is of $10^4$ K, whereas for the hot gas we test two different values, $5~\times~10^6$ K, and $10^7$ K. 
In most of the simulations without star formation, the effects of gravity are neglected. 
This is based on the fact that in B18, from the calculation of the virial ratio of the cloud,
it was shown that the system is not bound gravitationally.
In this paper, we investigate this aspect further and run one 
simulation in which the self-gravity of the gas is taken into acccount,
with the aim of testing its effects on the motion of the gas cloud in absence of star formation.
All the simulations without SF have been run for 1 Gyr. 
Finally, we will also present one simulation in which the effects 
of star formation and stellar feedback will be included. 

In Tab.~\ref{tab_models}, columns 9-11, we present a few global quantities computed from different simulations.
The cloud half-mass radius $R_{c, hm}$ (column 9 of Tab.~\ref{tab_models}) is computed from the two-dimensional mass density profile of
the cold (i.e. with temperature $T<20000$ K) gas. 
In each cell, the projected 2D density profile is computed in the y-z plane and with respect to an origin placed
at initial centre of the box. 
In all models but the Supersonic ones, initially the centre of the cloud coincides with the centre of the box.
In the Supersonic models, initially the cloud is at x=-10 kpc, y=0 and z=0. This choice is done in order to delay the leakage of
gas from the boundaries lying in the back of the cloud, which will occurr at earlier times in case of supersonic
velocity. 

In column 10 of Tab.~\ref{tab_models} we report the estimated line width $\sigma_l$. 
This quantity was computed from the 1D density-weighted velocity dispersion of the cold gas 
\begin{equation}
\sigma_{1D}^2 = \frac{1}{3}\frac{\Sigma \rho_c [(v_x - \bar{v_x})^2 + (v_y - \bar{v_y})^2 + (v_z - \bar{v_z})^2]}{\Sigma \rho}
\end{equation}
(e. g., \citealt{she10}), where $\rho_c$, $v_x$, $v_y$ and $v_z$ are the density, 
the x-, y- and z-component of the velocity of the cold  gas 
in a cell, respectively, whereas $\bar{v_x}$, $\bar{v_y}$ and $\bar{v_z}$ are the average x-, y- and z-component velocity values, respectively.

Beside $\sigma_{1D}$, the observed line width will include also a contribution
from the sound speed $c_s$: 
\begin{equation}
\sigma_l^2 = \sigma_{1D}^2 + c_s^2 
\label{sigmal}
\end{equation}
The minumum temperature value adopted in this work  
is $5000$ K, hence the sound speed in the cold gas as defined here 
will range between 8 km~s$^{-1}$ and 16 km~s$^{-1}$. This implies that, for the temperature range
considered here, $\sigma_l$ will always be higher than the observed line width, for which we consider the value of $\sigma_{\rm HI}=9$ km/s
reported in Tab.~\ref{tab_obs} and which is very likely to trace gas with T$<5000$ K and lower sound speed.
For these reasons, in Tab.~\ref{tab_models} for the line width we provide upper limit values computed assuming in
Eq.~\ref{sigmal}  
$c_s$ = 16 km~s$^{-1}$, which correspond to the maximum temperature value in our definition of {\it cold gas}, i.e. T=$20000$ K. \\
Finally, in the last column of Tab.~\ref{tab_models} we report the average star formation value for the simulations that include
star formation and feedback. All the quantities presented in columns 9-11 have been computed from the results of the runs
obtained at epochs between 0.25 Gyr and 1 Gyr.

\subsection{Cooling and heating}
\label{sec_thermal}
Radiative cooling of the gas is computed assuming photoionization equilibrium
as a function of temperature for different values of the gas density and metallicity. 
The cooling function implemented in RAMSES takes into 
account H, He and metal cooling (see \citealt{few14}).
At the beginning of our simulations the metallicities of the cloud
and of the ICM are uniform, but during the evolution of the systems metals are advected,
and can also be newly produced if new stars are born.
At each timestep and in each cell 
the metallicity value is stored and continuously updated into a passive tracer,  
which is used by the radiative cooling routine 
to take into account the evolution of the metal content of the gas.\\

The contribution from metals is accounted for by means of a 
fit of the difference between the cooling rates calculated at solar metallicity and those at zero
metallicity using the photoionisation code
CLOUDY (\citealt{fer98}).
The temperature threshold for radiative cooling adopted in this work is $5000$ K,
i. e. anywhere in the volume the gas is not allowed to cool down to a temperature below this value. 
The aim of this choice is to model a two-phase medium, consisting of 
a neutral one, representing the coldest regions of the cloud, and an ionised one which includes the ICM and
the gas continuously exchanged between the cloud and the external medim. \\
In our simulations, we do not take into account thermal conduction. \\
The effects of a uniform ultaviolet (UV) background are also taken into account. 
The RAMSES cooling routine includes functional fits for the photo-heating and photoionisation rates of the \cite{haa96}
ionizing
background spectrum,  
as formulated in \cite{the98}. \\
Besides radiative cooling and heating from a UV background, in our star formation
simulation we also include stellar feedback as further heating source (see Sect. \ref{sec_sf}).

\subsection{Star formation and stellar feedback}
\label{sec_sf}
The star formation (SF) model used in this work relies mostly upon the native
RAMSES implementation, which is described in detail in \citealt{ras06}.
In each cell where the flow is converging, 
we assume that the gas with temperature $T<2 ~ 10^4$ K can form one star
particle. In each cell where these criteria are met, 
the gas, characterised by density $\rho$, can be converted into star particles with density ${\rho}_{*}$ according to:
\begin{equation}
  \dot{\rho}_{*} = \frac{\rho}{t_*}, 
\end{equation}
i.e. according to the \cite{sch59} law.

The star formation timescale $t_*$ is proportional to the local free-fall time $t_{ff}$, and computed as
\begin{equation}
  t_{*}=  t_{ff}/\epsilon_{ff},  
\end{equation}
where $t_{ff}=\sqrt{3~\pi/32~G \rho}$ and $\epsilon_{ff}$ is the star formation efficiency per free-fall time. 
As described later in Sect.~\ref{sec_results},  
various values for $\epsilon_{ff}$  have been tested in lower resolution
simulations, in order to account for some observational features of SECCO~1.\\
For purposes of numerical stability, the code allows that no more that 90\% of the gas in
a cell can be used for star formation.  
Star particles are spawned stochastically as detailed below. 
The mass of each collisionless stellar particle is $M_p= N~m_0$, where $m_0=25~M_{\odot}$ is the minimum particle mass
and $N$ is determined by sampling from a Poisson distribution, characterised by a probability $P$ given by
\begin{equation}
P(N)={\lambda_P \over N !} \exp({-\lambda_P}) \, ,
\label{poisson_law}
\end{equation} 
where the mean value is calculated as: 
\begin{equation}
\lambda_P= \left ( {\rho \Delta x^3 \over m_0}\right ) {\Delta t \over t_*}. 
\label{poisson_param}
\end{equation}
In this equation $\Delta t$ represents the timestep, whereas $\Delta x$ is the cell size.

The formation of single stars cannot be resolved in our simulations, hence one
star particle is meant to describe a stellar association or a group of stars. 
The collisionless star particles are placed at the centre of their parent cell,
with a velocity equal to the local fluid velocity.
The corresponding mass, momentum, energy and density  
are conservatively removed from the ones of the parent cell.
Once formed, star particles are allowed to move as an N-Body particles in the
gravitational potential of the system, which is computed from its total mass. 
Each star particle can inject energy, mass and metals into the surrounding gas.
Once born, each particle immediately deposits in the cell an amount of
mass equal
to $\eta~M_p$, with $\eta=0.1$, and an amount of energy
\begin{equation}
E_p = \frac{\eta~M_p}{10~M_\odot} \cdot 10^{51} erg,  
\end{equation} 
as well as a mass of metals which can be computed directly from its initial mass
and the metallicity of the gas out of which it has originated. 
The ejection of $\sim~10~\%$ of the initial mass of stellar particles
is expected if each of them represents a simple stellar population
with a standard (e. g., \citealt{sal55} or \citealt{kro01}) 
initial mass function, and if each massive star (i.e. the stars with mass $M>8~M_{\odot}$) 
ejects most of its mass during its lifetime, leaving behind a remnant of a few solar masses. In the case in which SF and feeback have been taken into account, the simulation has been terminated at 2 Gyr.

\begin{figure*}
	\includegraphics[width=20.5cm]{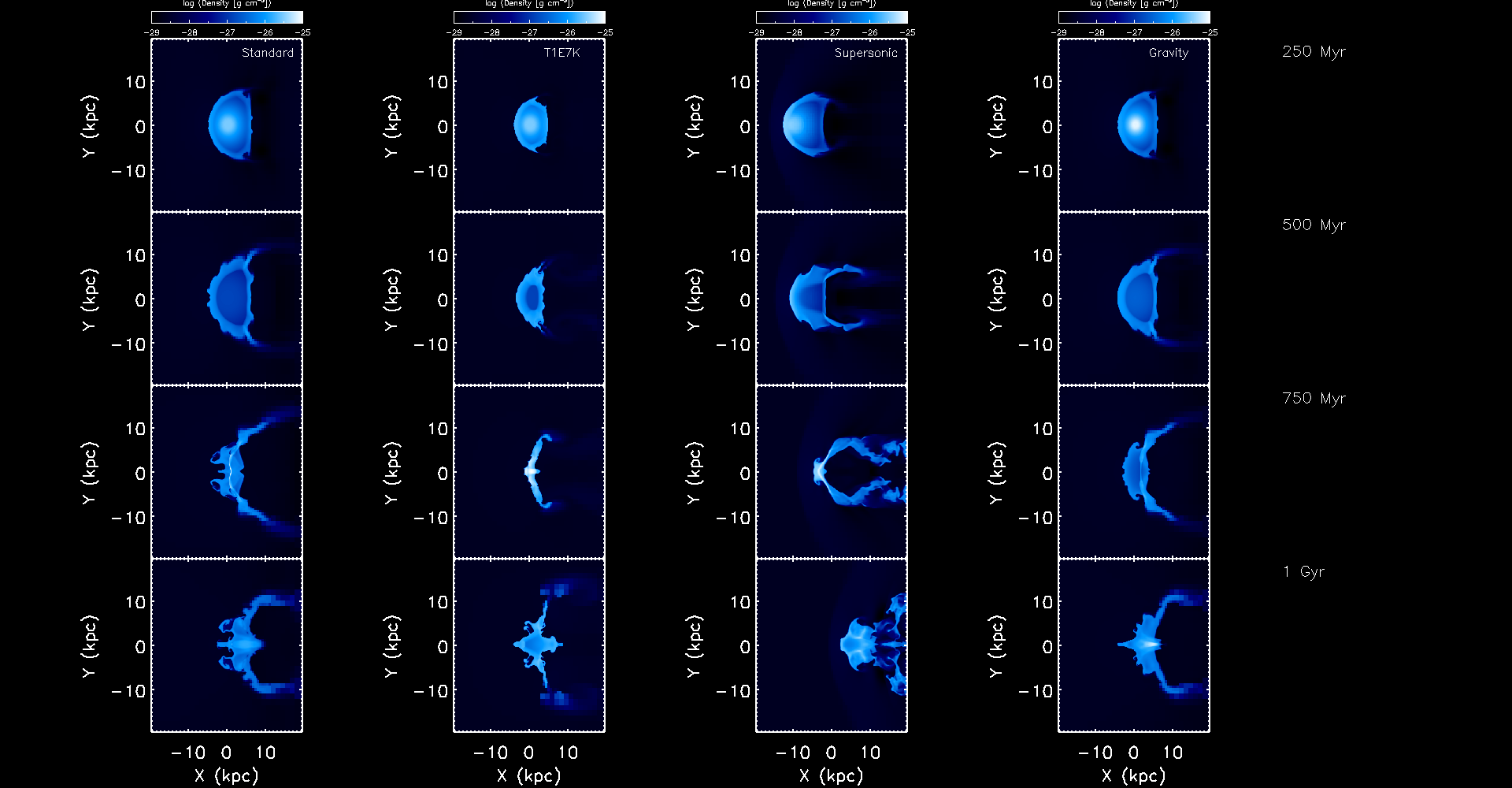}
    \caption{Two-dimensional gas density maps in the x-y plane in our simulations at various time steps
    for the Standard (first column), T1E7K (second column), supersonic (third column) and gravity (fourth column) simulations. 
    In each column, the first, second, third and fourth row describe the model at 250 Myr, 500 Myr, 750 Myr and 1 Gyr, respectively.    }
    \label{fig_dmap}
\end{figure*} 

\begin{figure*}
	\includegraphics[width=20.5cm]{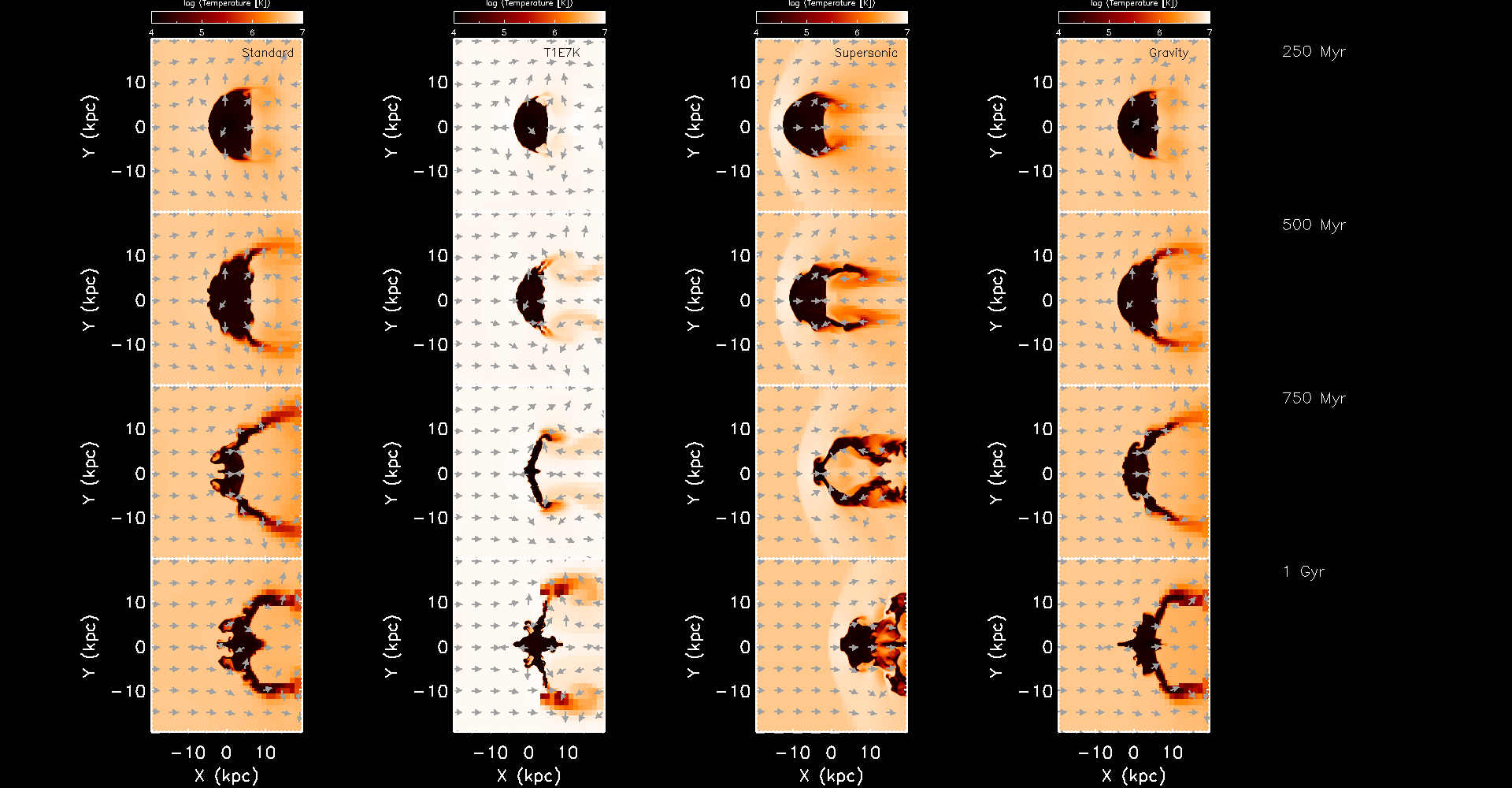}
    \caption{Two-dimensional gas temperature maps in the x-y plane in our simulations at various time steps
    for the Standard (first column), T1E7K (second column), supersonic (third column) and gravity (fourth column) simulations. 
    In each column, the first, second, third and fourth row describe the model at 250 Myr, 500 Myr, 750 Myr and 1 Gyr, respectively.
    In each panel, the gray arrows describe the velocity field.}
    \label{fig_tmap}
\end{figure*}

\begin{figure*}
\begin{center}
\includegraphics[width=3.8cm]{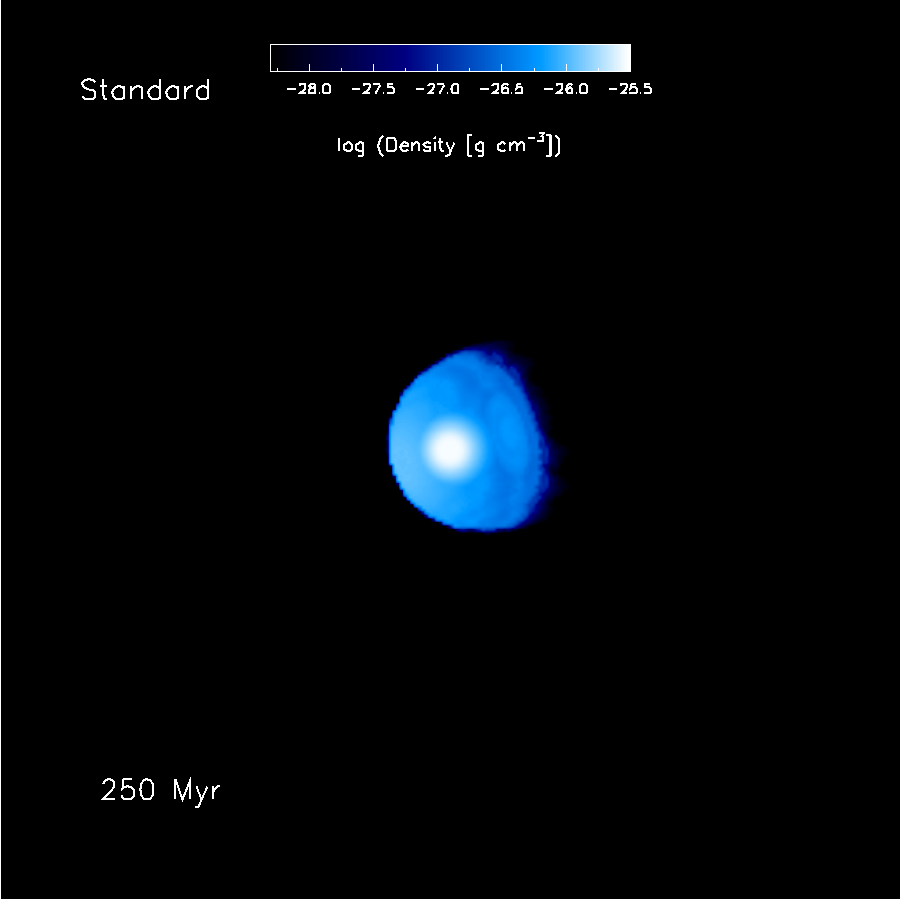}
\includegraphics[width=3.8cm]{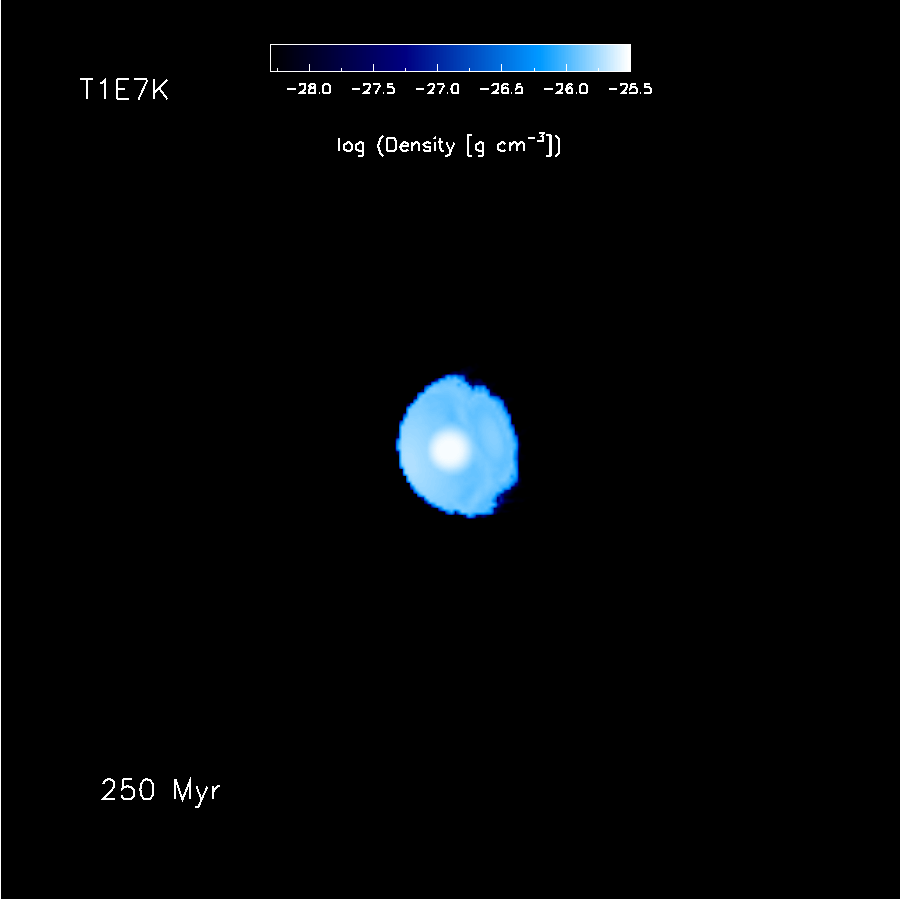}
\includegraphics[width=3.8cm]{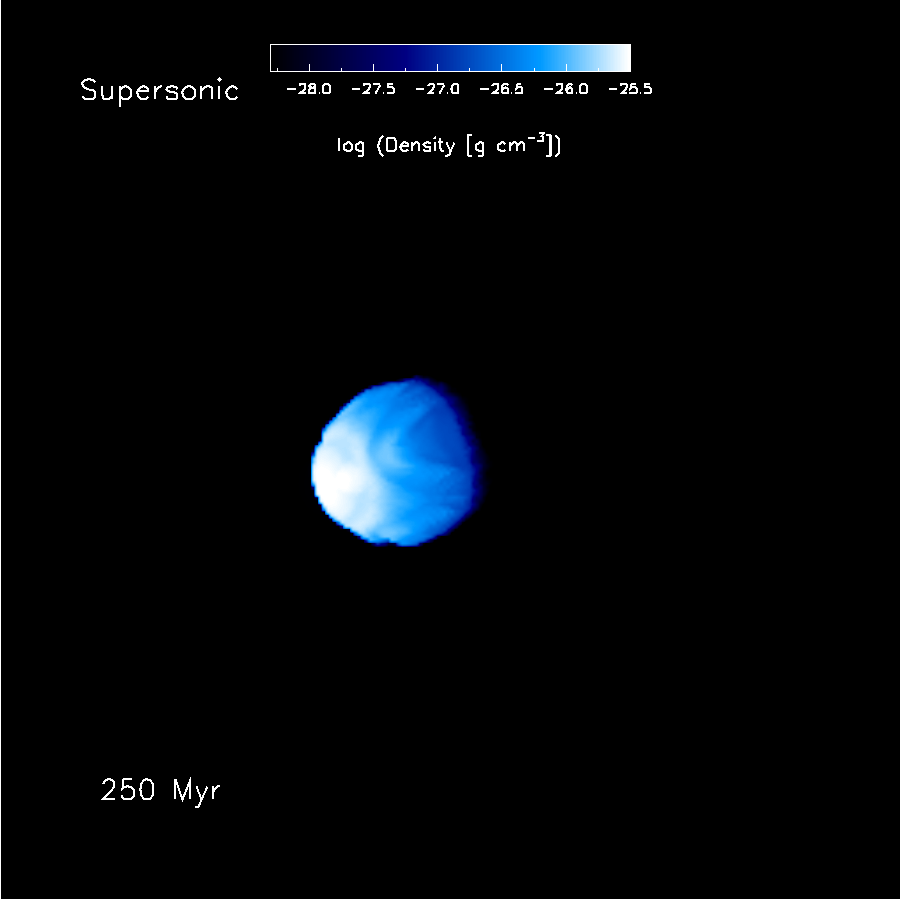}
\includegraphics[width=3.8cm]{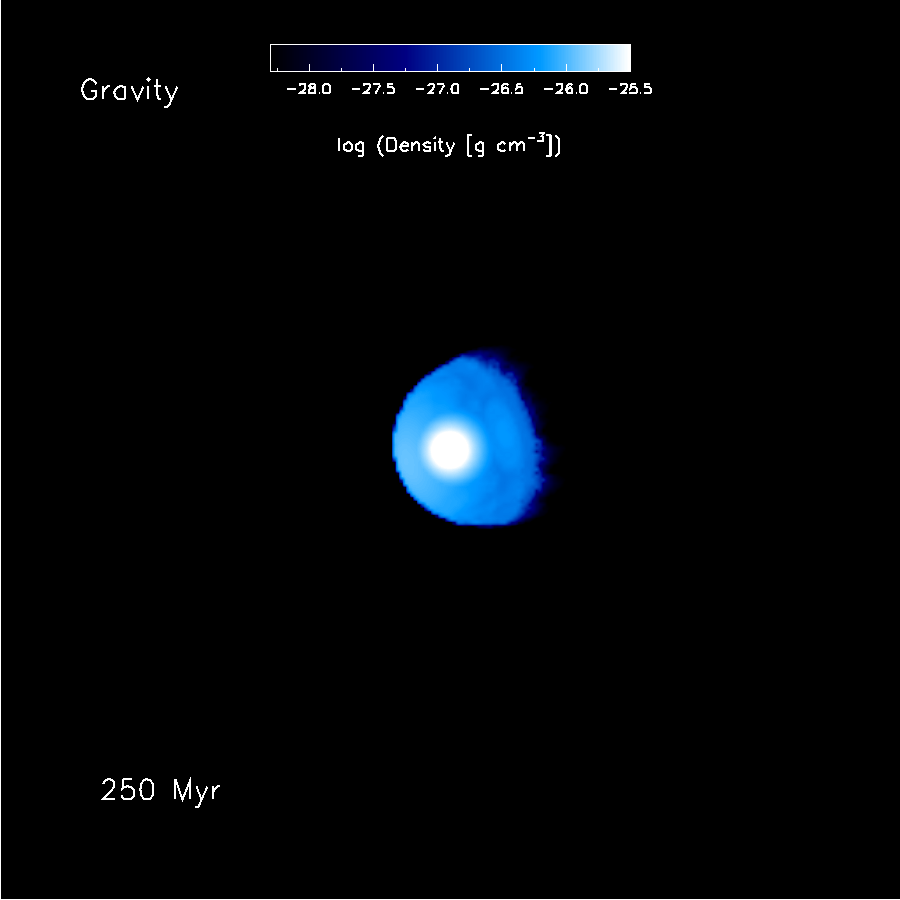}

\includegraphics[width=3.8cm]{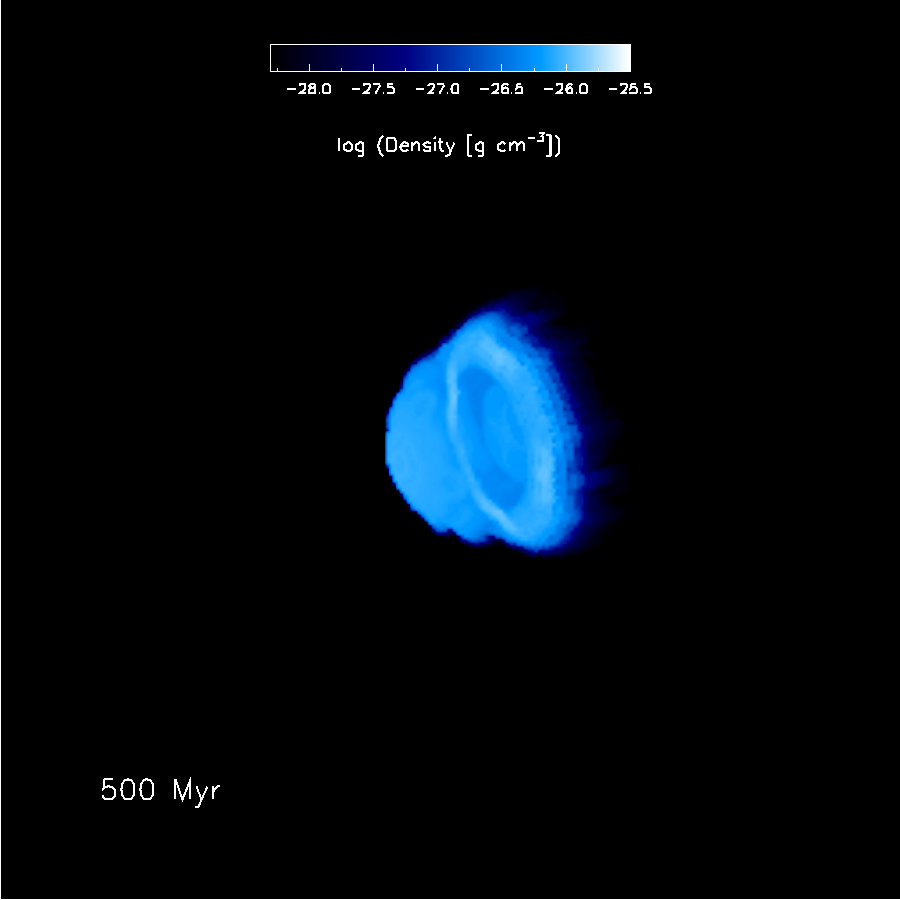}
\includegraphics[width=3.8cm]{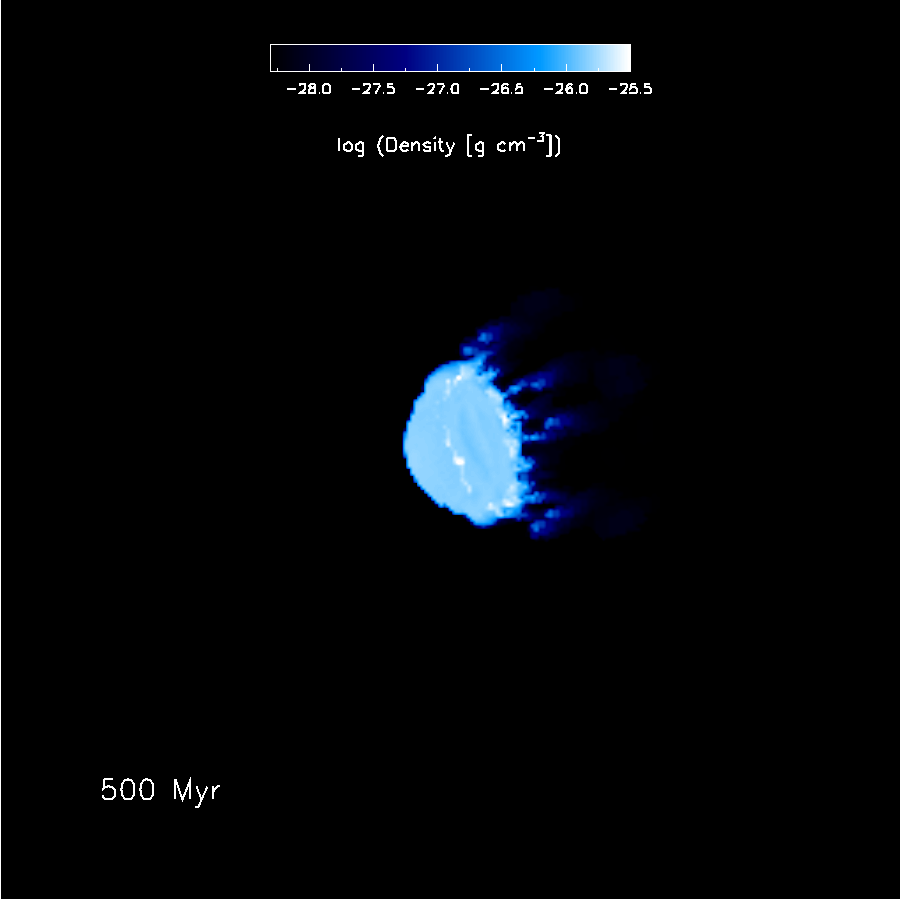}
\includegraphics[width=3.8cm]{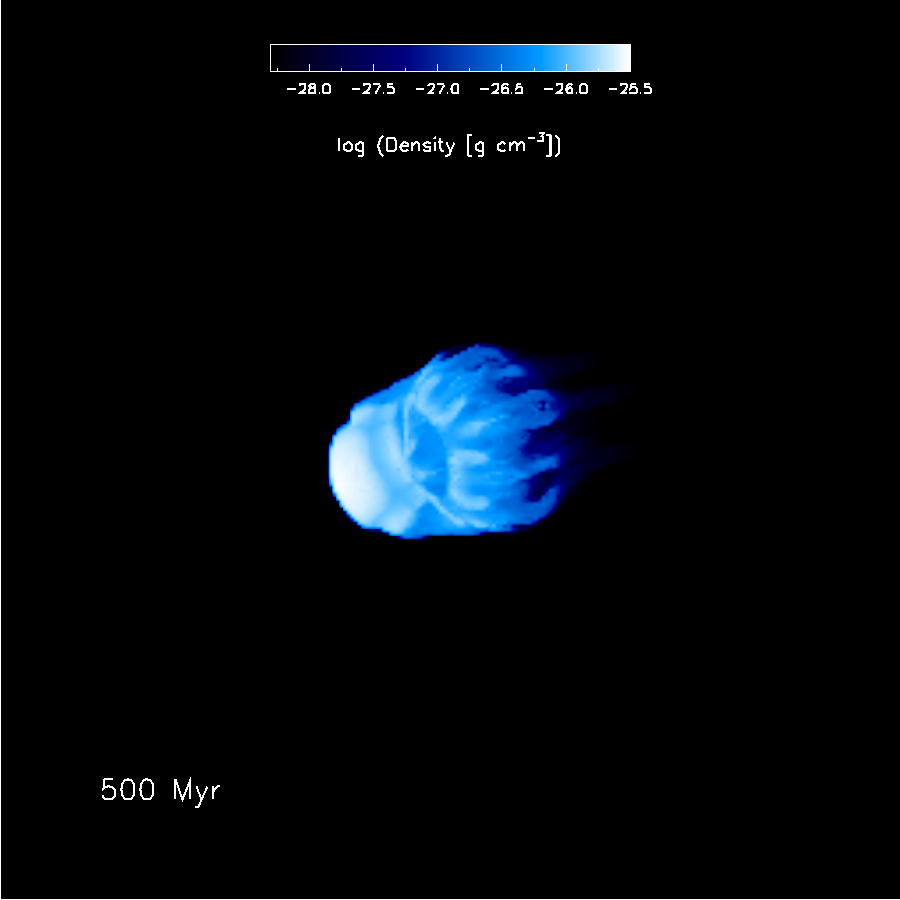}
\includegraphics[width=3.8cm]{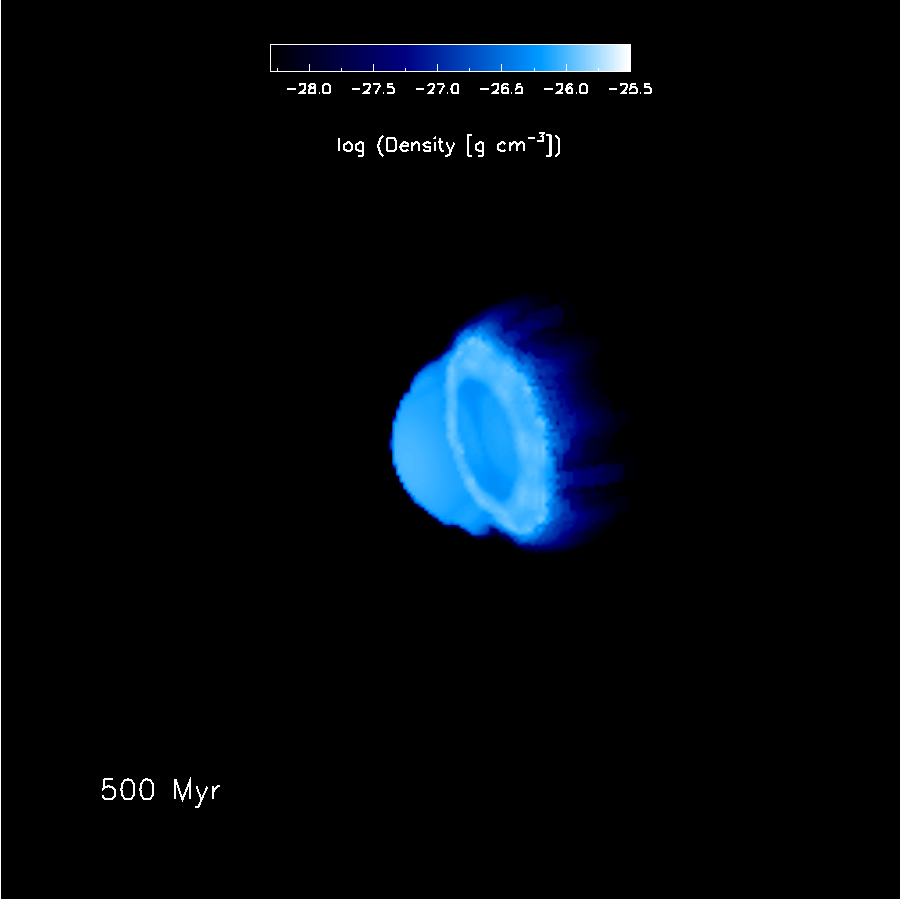}

\includegraphics[width=3.8cm]{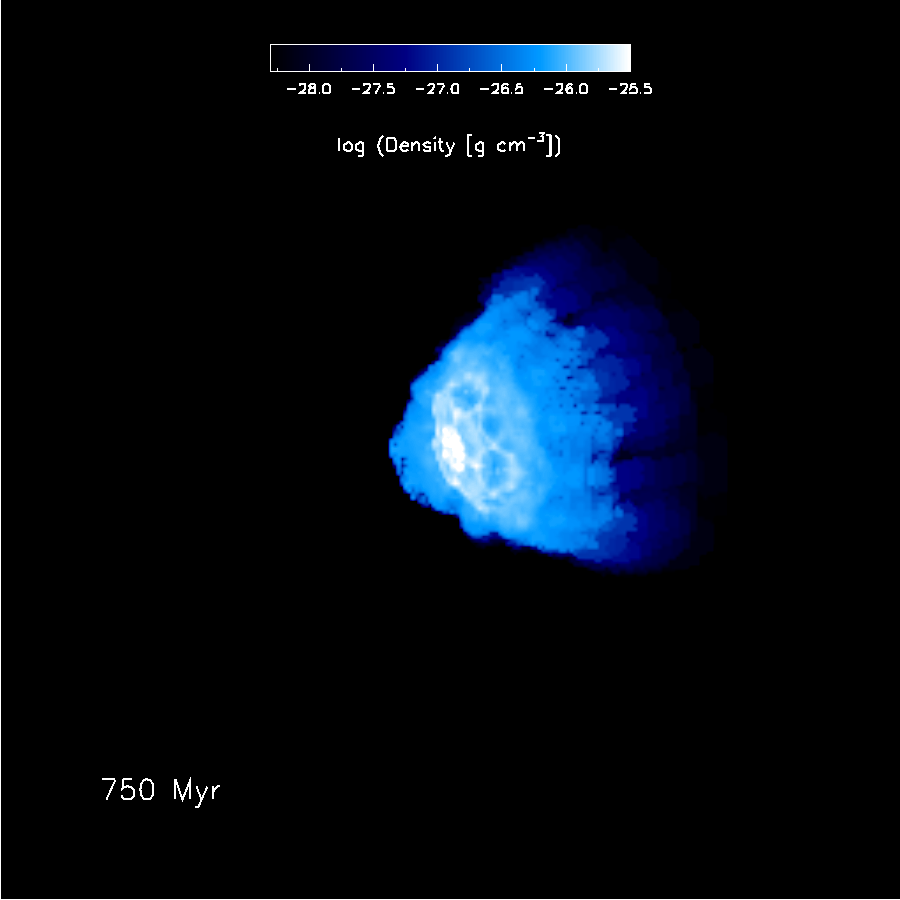}
\includegraphics[width=3.8cm]{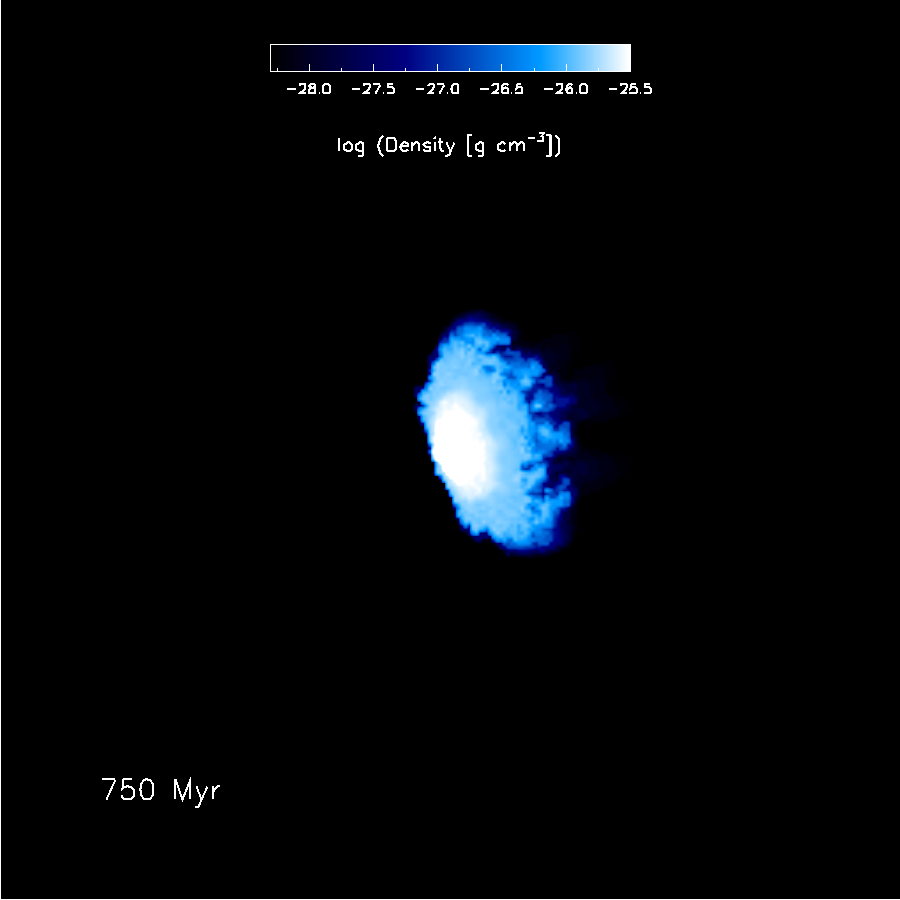}
\includegraphics[width=3.8cm]{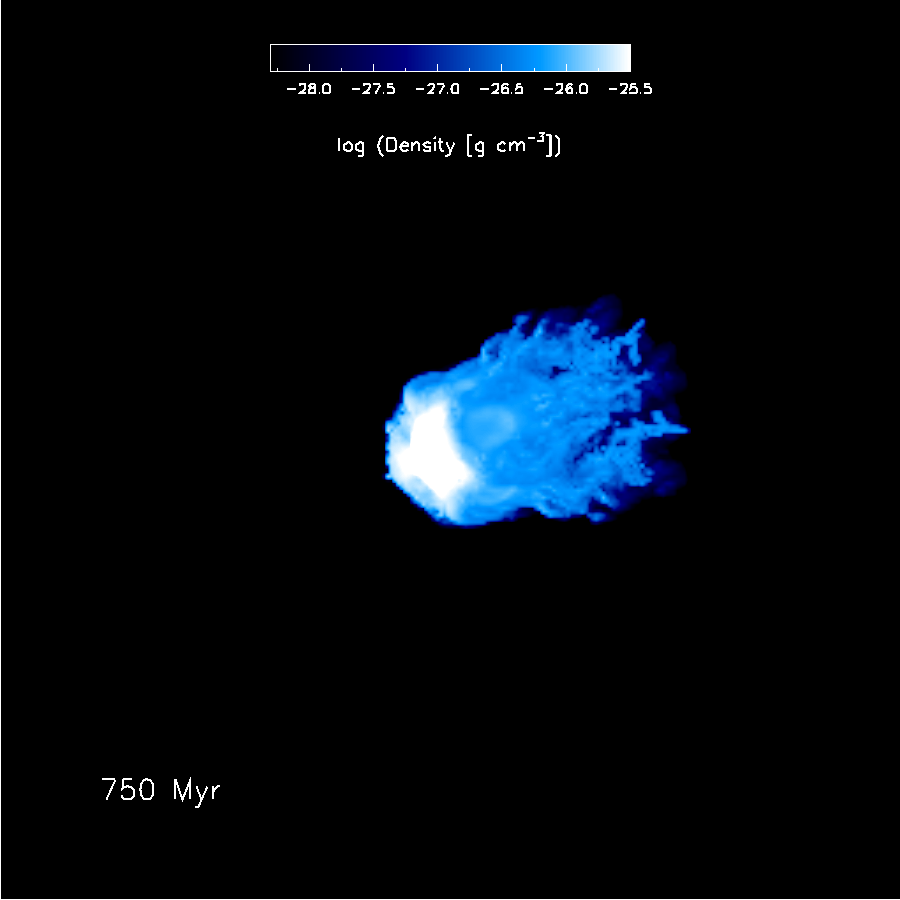}
\includegraphics[width=3.8cm]{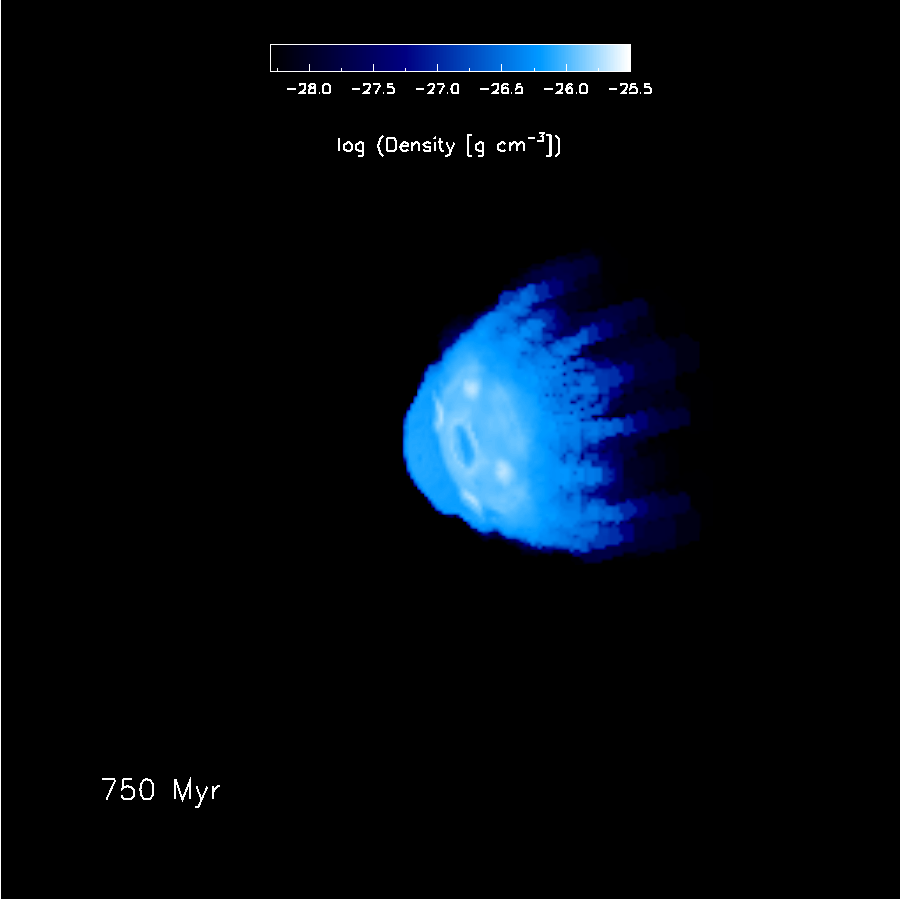}

\includegraphics[width=3.8cm]{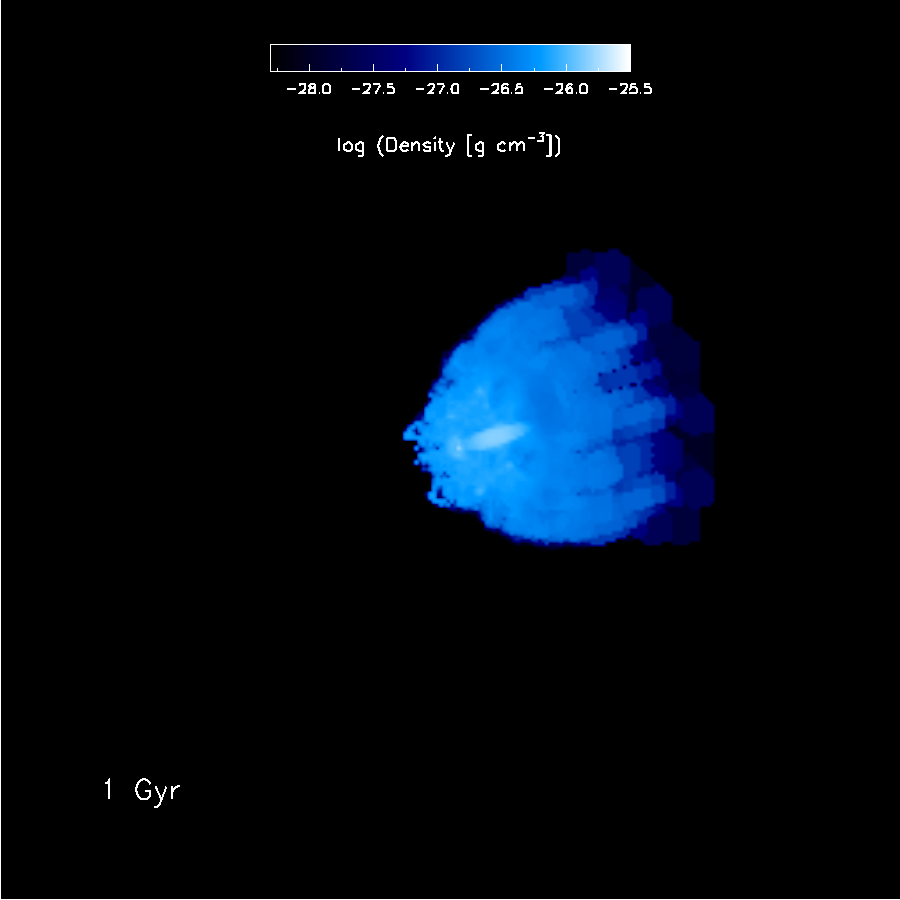}
\includegraphics[width=3.8cm]{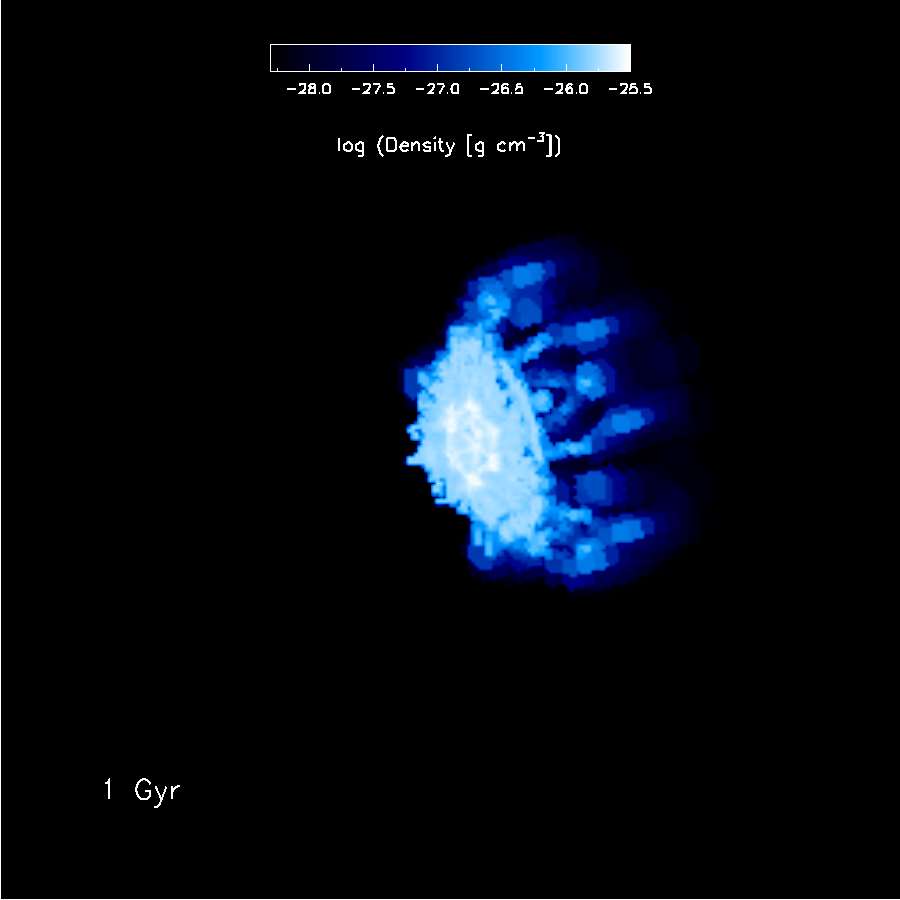}
\includegraphics[width=3.8cm]{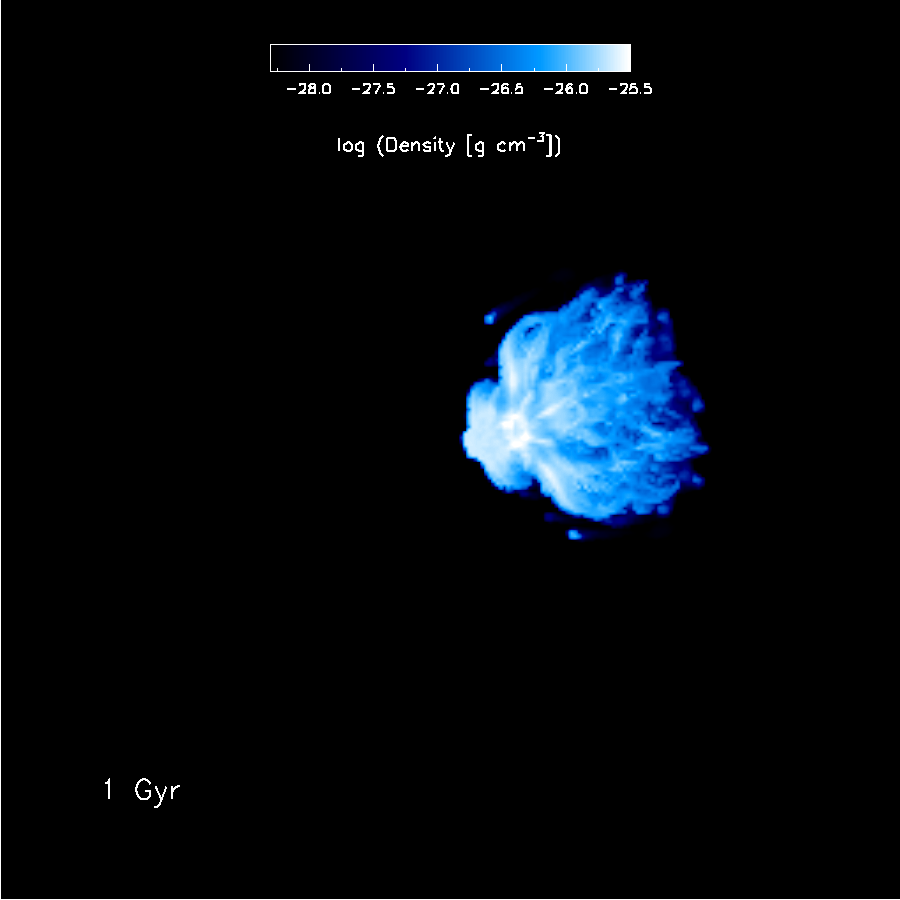}
\includegraphics[width=3.8cm]{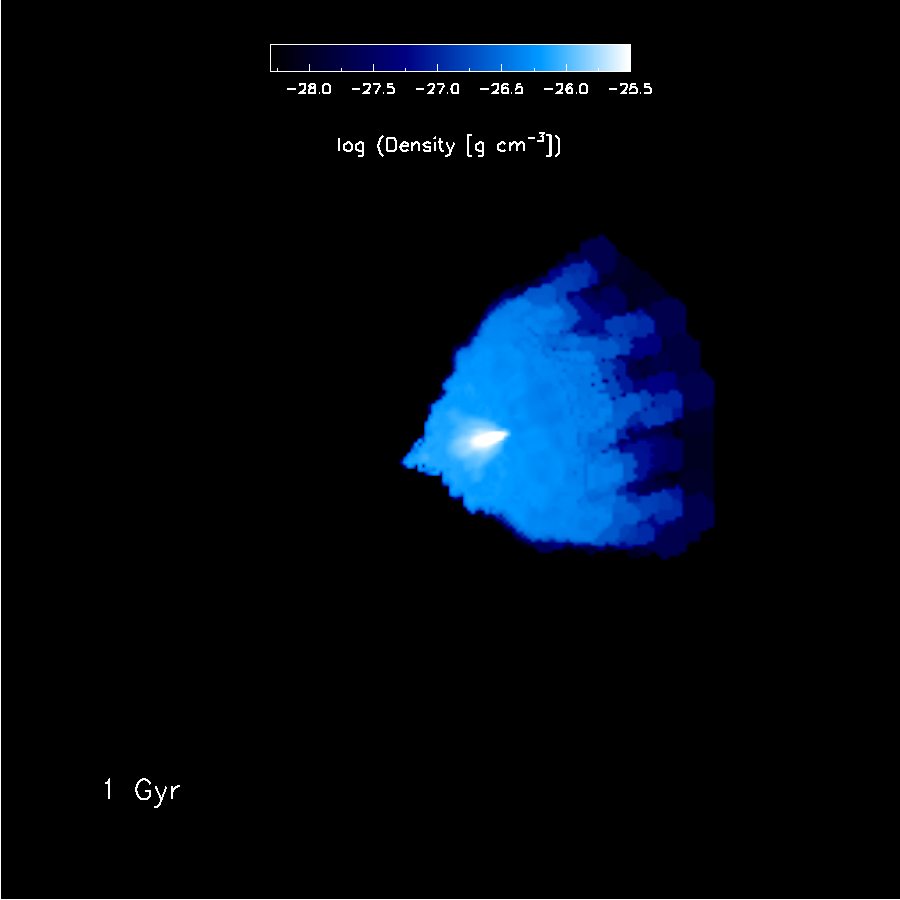}

\end{center}
\caption{Three-dimensional gas density distribution in our simulations at various time steps
    for the Standard (first column), T1E7K (second column), supersonic (third column) and gravity (fourth column) simulations. 
    In each column, the first, second, third and fourth row describe the model at 250 Myr, 500 Myr, 750 Myr and 1 Gyr, respectively.
    The assumed rotation angle with respect to the x- and y- axis is of 30 degrees.}
    \label{fig_tdmap}
\end{figure*}

\begin{figure*}
        \includegraphics[width=18.5cm]{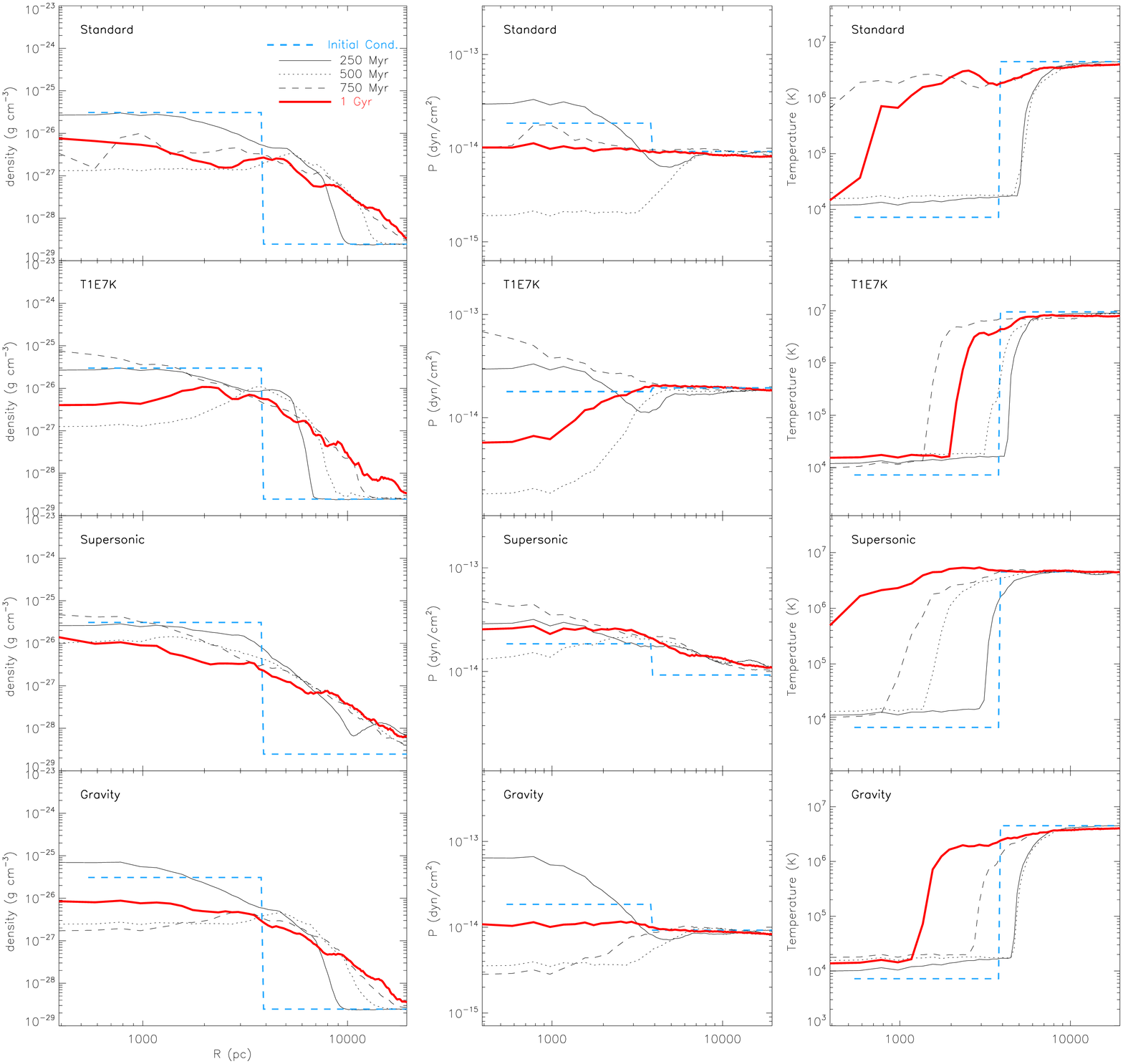}
    \caption{Density (first column), Pressure (second column) and temperature (third column) profiles computed
    for our simulations at different timesteps. The first, second, third and fourth row include the profiles of
    the Standard, T1E7K, supersonic and gravity models, respectively.
    In each panel, the blue thick dahed line describes the initial conditions of the Standard model, the thin solid, dotted and dashed lines are the
    profiles computed at 250 Myr, 500 Myr, 750 Myr, respectively, and the solid thick, red line is the final profile, computed at 
    1 Gyr. 
    }
    \label{fig_prof}
\end{figure*}


\section{Results}  \label{sec_results}
The origin of SECCO~1 and, in particular, the candidate parent 
galaxies have been thoroughly discussed in B18.
In all the considered cases, a travel time $\sim 1$~Gyr through the Virgo ICM 
is required for SECCO~1 to
reach its current position.
Based on this argument, as in B18, most simulations are run for at least 1 Gyr, 
to study if the system can have retained its cold gas and 
survived within this  environment for such a long time.\\
The stability of the gas cloud and the general behaviour
of the system is studied at various evolutionary  
epochs by means of various diagnostics, i.e. two-dimensional density and temperature maps,
radial profiles for density, temperature and pressure and three-dimensional gas distribution.
All the profiles discussed in this section are computed by assuming an origin placed at the centre of the computational box. 
The temporal checkpoints for our analysis are 0.25, 0.5, 0.75 and 1 Gyr, which
are the epochs at which the features of the sytems are studied and discussed.

\subsection{Hydrodynamical Evolution of the gas cloud}  \label{sec_nosf}
Figures~\ref{fig_dmap} and ~\ref{fig_tmap} show two-dimensional density and temperature maps, repectively,
computed at four different evolutionary times for four of the models described
in Sect.~\ref{sec_model}. The models we consider in this section
are the 'Standard' one (in which the initial parameters are the same as the ones of B18),
the T1E7K, the Supersonic and the Gravity ones, 
i. e. all the models in which star formation and stellar feedback within the cloud
are not taken into account. 
Each map represents a section of the computational domain in the x-y plane; 
the maps have been calculated by selecting all the grids crossed by a plane
centered at the middle of the computational volume and perpendicular to the {\it z-}axis.  
In all the cases shown in Figures ~\ref{fig_dmap} and ~\ref{fig_tmap},  the evolution of the system 
is mosly driven by the different degrees of compression attained in various cases,
and by the growth of Kelvin-Helmholtz instabilities.
To complement our description of the physical properties of the cloud in these four cases as
visible from the 2D maps, for each one of them we also show the 3D volume-rendering maps (see \citealt{pom08}) 
of the gas density
at different times (Fig.~\ref{fig_tdmap}) and the evolution of the density, pressure and
temperature profiles (Fig.~\ref{fig_prof}).

\subsubsection{The Standard model}
The first columns from the left of of Figures ~\ref{fig_dmap} and ~\ref{fig_tmap} show the time evolution
of the density and temperature, respectively, of the cold gas cloud in the x-y plane
in the case of the Standard model,
for which the initial setup is the same of the 
3D simulation set described in B18. 

As in the 3D model of B18, even if an overall stability of the cloud was
found across a time interval of 0.25 Gyr, already at this time 
the cloud presents a shape significantly different than the intial, spherical one, 
visible from various aspects. 
First of all, a compression in the back of the cloud causes a flattening of its rightmost edge
and a central overdensity, as visible also from the density profile in Fig.~\ref{fig_prof}. 

The shape of the cloud also changes significantly because of the Kelvin-Helmholtz or
shear instabilities, which are caused by the velocity difference between the cloud and incoming fast hot intra-cluster gas. 
Symmetric, tiny filaments are visible at  
the upper and lower edges of the cloud and at the extremes of its back. 
As visible from the temperature map (Fig.~\ref{fig_tmap}), such
filaments are surrounded by gas layers which have intermediate temperature between the gas cloud and the
ICM.

The velocity field (gray arrows in Fig.~\ref{fig_tmap})
shows that at large distances from the centre the motion of the incoming gas is
weakly perturbed by the presence of the cloud, with nearly parallel arrows
in the leftmost side of the plot and at y-distances $>10$ kpc.
On the left side of the temperature map (i. e. at negative values of the x-coordinate), 
the arrows are tilted by the rise of a few parcel of instable cold gas, 
which from the cloud protrude outwards. 
In proximity of the cloud, the inclination of the arrows is gradually increasing along the y-axis,
to become parallel to the y-axis at Y$\sim 10$ kpc. Moreover, downstream of the cloud some of the hot gas
presents an inverted velocity field with respect to the overall flow.
This feature is visible from the inclination of the filaments, which increases
progressively as one moves towards higher values of both the x- and y- coordinates, to become parallel to
the x-axis at $|$Y$|~\sim 9$ kpc.
At the same value of the y-coordinate and at X $\sim 10$, the instability curls up and some of the gas moves
parallel to the flow but in opposite direction, 
which explains the negative x-component of the velocity. 
The curling-up of the extended unstable wings far from the cloud is barely visible in the maps also because of
the AMR, which uses low resolution at large distances from the cloud centre 
(where the density constrast is not significant), but more curled-up extended filaments and smaller rippled structures are
visible in other maps (e. g. in the temperature map of the T1E7K and supersonic models at 0.75 and 1 Gyr).

After 0.5 Gyr, the cloud has significantly expanded, the ripples which were barely visible at earlier times
are now much more prominent in both the front and rear profile and distributed in
a symmetric fashion with respect to the x-axis. 
The two overdense bent filaments at the two upper and lower sides 
have grown considerably, and now extend up to the righmost edge of the 2D maps.
The velocity field reveals a complex pattern, 
with parcels of cold gas which, from front of the cloud, expand forward
and with an inverse velocity with respect to the bulk flow of the ICM.  

A more elongated shape of the cloud is also visible in the 3D map, which shows that the
rear part has undergone more extension than the head. At this time, the density profile is flatter than
at 0.25 Gyr and overall, the cloud is still much colder and with a much lower pressure with respect to the external
environment (dotted lines in the three top panels of Fig.~\ref{fig_prof}).

At 0.75 Gyr, the x-y cross section of the cloud is significantly thinner than at earliest times.
Several instabilities which were before weakly prominent along the cloud profile
are now much more developed. 
Two thin filaments have emerged from the two instabilities at the sides of the front and they have
curled-up backwards, rejoining the main body of the cloud. 
A very dense and thin shell is visible at the centre of the cloud, extending along the y-axis for several kpc,
and which is visible in its entire extent in the 3D map, which reveals 
an intricate pattern of filaments in the central flat dense gas distribution. 
At this time, the density profile has flattened further. Thanks to its considerable expansion and
to the heating of a significant portion of the originally cold gas, 
the cloud has nearly reached pressure equilibrium with the external environment (dashed lines in 
in the three upper panels of Fig.~\ref{fig_prof}). 

At t=1 Gyr, the central parts have re-expanded with respect to the previous
timestep, in particular along the x-axis at Y$=0$. The structure still pertains its
symmetric shape. At this time, a complete pressure equilibrium with the hot ICM has been reached, vible from the flat pressure
profile at 1 Gyr (red thick solid  lines in the three top panels of Fig.~\ref{fig_prof}). 

The results described in this section have shown how a timescale of $\sim$0.5 Gyr is enough for a system such as
SECCO~1, to evolve from an initial spherical shape (as assumed here) to an umbrella-like or jellyfish-like shape.
In the Standard model, the cloud has typical size (as measured by its half-mass radius $R_{c, HM}$ in Tab.~\ref{tab_models})
between 3.3 and 6.4 kpc, where in general lower values are found at earlier times, and line with $\sigma_l < 20.8$ km/s,
which corresponds to a maximum 1D velocity dispersion $\sigma_{1D}=$4.8 km/s.

\subsubsection{The effects of different initial conditions on the evolution of the cloud}
\label{sec_nosf}
Some of the values assumed for a few fundamental initial parameters of the SECCO~1 Stardard simulation 
are clearly uncertain. 
For instance, the temperature assumed for the external medium is of $\sim~5~10^6$ K,
but considering the temperature 
profile of the Virgo cluster (\citealt{urb11}) and that SECCO~1 could lie at a distance of $\sim 1.1$ Mpc from the centre of the cluster,
in principle the ICM temperature could also be higher, of the order of $10^7$ K.\\
Moreover, in B18 a velocity of $200$ km/s was assumed from the estimated projected value. 
It may be interesting to study the effects of a larger velocity,
and in particular of a supersonic motion, on its stability. Also gravity could influence its evolution, in particular
by increasing its stability against the development of filamentary structures or instabilities. 
In this Section we discuss how different physical assumptions regarding the initial conditions and some main parameters of the simulation 
can affect the long-term evolution of the system.

{\bf i. Effects of a larger ICM temperature value}. The second columns from the left of of Figures ~\ref{fig_dmap} and ~\ref{fig_tmap} show the time evolution
of the x-y density and temperature maps, respectively, of the gas cloud 
when a larger temperature value of $10^7$ K is assumed for the Virgo ICM gas.
In this case, at the beginning of the simulation the 
cloud is in a state of quasi-equilibrium of pressure with the ICM (blue thick dashed lines the the second row of Fig. 4). 
At early times, the cloud appears much more compressed than in the Standard model.
This is clearly visible from its size at 0.25 Gyr as shown in the density map.
In the innermost regions, the density profile is not much different than the one of the Standard model
(solid lines in the second row from top in Fig.~\ref{fig_prof}),
but in this case it is truncated at $R~\sim 5$ kpc, whereas with lower ICM temperature it decreases more
slowly up to $R~\sim 10$ kpc.
Despite the different aspect, the instabilities at the edges of the cloud are still visible also in the case of a larger ICM temperature.\\ 
From the pressure profile, we can see how 0.25 Gyr of evolution are enough for the cloud to move out of equlibrium of pressure with the ICM.\\ 
The stronger compression of the cloud with respect to the Standard model is more visible at 0.5 Gyr.
At this time,  both the density and pressure in the cloud have increased with respect to the previous epoch
(dashed lines in Fig.~\ref{fig_prof}), at variance with the Standard model in which these two
quantities generally decrease at later epochs. 
Several filaments behind the cloud are visible in the 3D rendered map in Fig.~\ref{fig_tdmap},
as well as in Fig.~\ref{fig_tmap}.\\
At a later time (0.75 Gyr), the x-y cross-section of the cloud is considerably denser and thinner than the one
of the Standard model (and the thinnest of all the cases considered here). 
Due to the strong compression in all directions, the 3D structure of the cloud appears considerably flat (Fig.~\ref{fig_tdmap}). \\
A thinner x-y cross section with respect to the Standard model 
and a flat shape of the cloud is visible also later at 1 Gyr, although 
significant expansion has occurred at this stage. 
At variance with the Standard model, one striking feature of the system emerging from the pressure profile is that,  
after 1 Gyr,  it has not reached pressure equilibrium with the external environment.
Although the pressure profile increases smoothly outwards,  
the pressure value in the central regions of the cloud is by a factor $\sim3$ lower than the one of the ICM. 

Fig. ~\ref{fig_cold} shows the time evolution of the cold gas fraction in the cloud in the Standard model and
in the other cases discussed here.  
As in B18, the cold gas fraction has been computed considering the gas with temperature $<20000$ K.
At a given time, the gas fraction is the ratio between the total mass of gas with temperature below such value,
divided by the initial mass of cold gas, i. e. by the initial mass of the cloud.
Overall, the assumption of a high temperature in the ICM causes the system to retain larger fractions of cold gas
with respect to the Standard model. This occurs as a higher ICM pressure
causes a stronger contraction of the cloud, hence higher gas density and consequently a higher cooling rate,
which on the long term evolution of the cloud
facilitates the retention of the cold gas. 
Assuming a larger ICM temperature affects also the size of the cloud, with maximum values of its half-mass radius of 4.2 kpc
(Tab.\ref{tab_models}). 

{\bf ii. Effects of a higher relative velocity}.
In the case of a supersonic relative velocity between the cloud and the ICM, initially the 
cloud has been placed closer to the leftmost boundary than in the Standard case.
The reason is that with a larger speed, the cloud undergoes a significantly larger expansion along the direction of motion.
Had not we adopted such an initial position, a considerable fraction of the cold gas would have leaked out
of the computational box after $\sim 0.5$ Gyr\footnote{In general, due to a strong expansion of the cloud which occurs in all cases, some of the gas
present initially flows out of the box. In all the cases in which star formation was not considered,
the leakage of gas out of the box starts at times comparable to $\sim 0.8$ Gyr and at 1 Gyr
the total mass lost as due to gas leakage is never larger than 10\% of the initial mass.}.
A slightly larger expansion of the back of the cloud and a larger curvature of its front profile with respect to the previous models 
is visible at 0.25 Gyr in the case of the Supersonic model in Figures~\ref{fig_dmap} and ~\ref{fig_tmap} (third column from left),
as also the presence of a bow shock upstream the cloud.
In  Fig. ~\ref{fig_tmap}, we can appreciate that the velocity field of the ICM gas behind the cloud but
in proximity (i. e. within a few kpc) of its upper and lower edges appears less distorted than in the Standard model.
The two main ripples at the upper and lower edges are now less developed.
The 2D density map and the 3D map (third column from left in Fig.~\ref{fig_tdmap})
emphasise the strong compression occurring at of the front of the cloud in the supersonic model.\\
At 0.5 Gyr, the cloud is still characterised by a central density comparable to the initial one
but by a larger pressure (third row from top in Fig.~\ref{fig_prof}).
A striking feature is the strong growth of two filaments in the back of the cloud as visibile in the x-y density map.
At a later time (0.75 Gyr), the cold filaments have curled up, their extent has reached the right boundary and they appear 
significiantly less expanded along the y-axis than in the Standard model.
The top of the cloud is still highly dense. 
The high velocity causes the baricenter of the cloud to shift along the x-axis and towards right. \\
At 1 Gyr, the elongated filaments have moved out of the computational box and the front of the cloud has expanded. 
The overall expansion of the cloud as traced by its 3D map is smaller than the one ofthe previous two models.
The pressure profile gently decreases 
from the central regions of the cloud to the outermost regions of the computational volume. \\
Fig. ~\ref{fig_cold} evidences that even in the supersonic case, the decrease of the cold gas fraction as a function of time is
lower than in the Standard model, and the general behaviour of this quantity is similar to the one of
the T1E7K model. One conclusion of our study is that both a large ICM temperature and a large velocity of the cloud
help increasing its capability to retain its initial cold gas reservoir across a 1 Gyr period.
The sharp decrease of the cold gas fraction of the Supersonic model at $\sim~0.9$ Gyr is partly due to effective loss of cold gas
from the simulation volume, as we have seen that at this time its cold filaments
have already reached one of the boundaries and some of the cold gas is leaking out of the box.
The assumption of a supersonic motion affects the size of the cloud as derived from its 
2D profile in a plane perpendicular to the x-axis, with maximum values of its half-mass radius of 4.4 kpc, i. e. lower than in the Standard model,
and causes line width values larger than in the Standard case, $\sigma_l< 22.6$ km/s, corresponding to maximum $\sigma_{1D}=$6.6 km/s.

In principle, the assumption of a  velocity $>400~$ km/s will lead to a faster cloud crushing. 
In adiabatic conditions, this can be understood by considering that 
the cloud crushing timescale scales with the inverse of the relative velocity between the hot medium
and the cloud \citep{kle94}. 
In a non-star forming cloud, radiative cooling can delay this disruption by a few cloud crushing timescales,
but not prevent it (\citealt{sca15} and references therein). \\
The effects of supersonic motion on a star-forming cloud will be discussed later in Sect.~\ref{sec_ssf}.

{\bf iii. Effects of the self-gravity of the cloud}.
In the simulations pesented in B18 and run to assess the stability of the cloud during its motion,
the self-gravity was neglected.
The reason was the low density of the cloud which, as discussed in B18, implies that
the system is unbound, stable against collapse and pressure-confined by the external medium.
As mentioned in B18, the global effect expected by gravity on the cloud is to limit 
the growth of hydrodynamical instabilities (see also \citealt{mur93}).\\
To directly assess the role of gravity in the evolution of SECCO~1, we run one simulation with same
initial conditions as in the Standard Model but also taking into account the self-gravity of the gas.\\
In presence of gravity, the two-dimensional density map indicates that at 0.25 Gyr the central density of the cloud is larger than
the one of the Standard model. The cloud also appears slightly more compressed than in the Standard model. 
An increased central density at this time is indicated also by the density profile shown in Fig.~\ref{fig_prof}. 
At this time, the 3D map computed for this model is barely distinguishable from the one of the Standard model.\\
At 0.5 Gyr the front profile of the cloud appears considerably smoother than the one of the Standard model.
A few instabilities which cause symmetric ripples on the front surface are absent,
whereas the two symmetric filaments lying behind the cloud and which extend backwards are still present, with
an aspect and size very similar to the one of the Standard model.
Also in presence of gravity the cloud has expanded by an amount similar to our Standard simulation,
which further supports the argument that gravity does not play a dominant role in its large-scale evolution.\\
At 0.75 Gyr, the prominent, curled up instabilities present on the front of the cloud in the 
Standard model are now absent, but the filaments on the back have reached out to the boundary.
Also the 3D figure shows that  
the front surface is overall smoother than the one of the Standard model, but also that 
the back instabilities do not seem to be underdeveloped. \\
At 1 Gyr the only frontal instability which has grown is the bar at the centre of the cloud. 
A sharp structure at the cloud front is evidenced also by the 3D map. The cloud is in pressure equilibrium with the ICM. \\
Fig.~\ref{fig_cold} shows how globally, gravity helps further in retaining the cold gas at all times, with differences in the gas fraction 
with respect to the Standard model which are less than $10\%$. \\
One conclusion of our study is that gravity hinders the development of small-scale instabilities 
all over the surface of the cloud, in particular on its front, but it does not prevent the growth of large scale
instabilities such as the elongated filament along its trail. 
The addition of self-gravity causes smaller cloud size values than in the Standard model, with $R_{c, HM}$ between 2.7 kpc and 5.8 kpc.

\begin{figure*}
        \includegraphics[width=10.5cm]{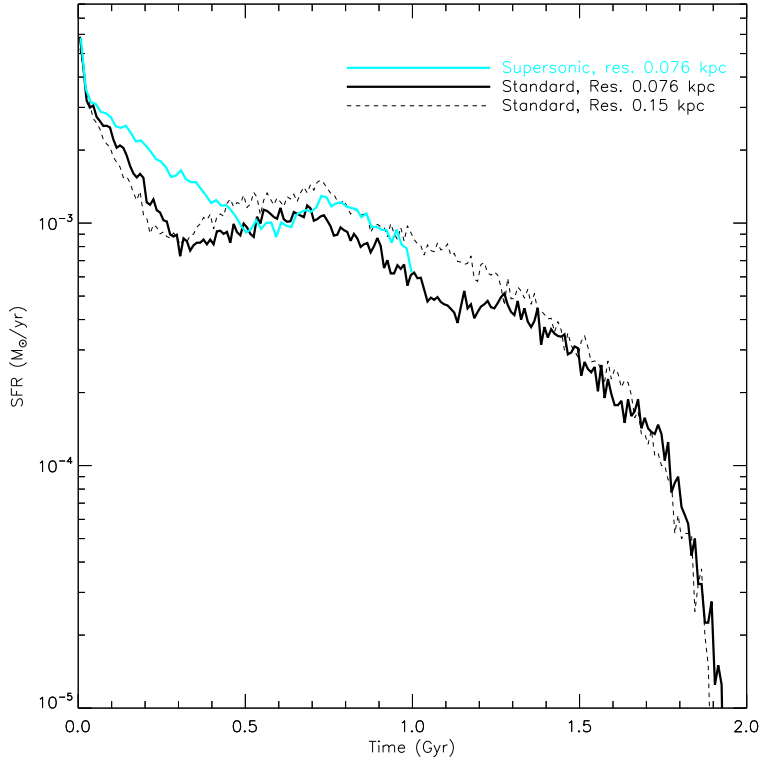}
    \caption{Star formation history of our Standard and Supersonic simulations. As for the Standard model, the black thick solid line is the SFH computed in our
    maximum resolution simulation, whereas the thin dashed line has been computed at lower resolution. The thick solid cyan line is the SFH of the Supersonic simulation
    at maximum resolution. In the latter case, the simulation has been stopped at 1 Gyr. }
    \label{fig_sfr}
\end{figure*}

\begin{figure*}
	\includegraphics[width=10.5cm]{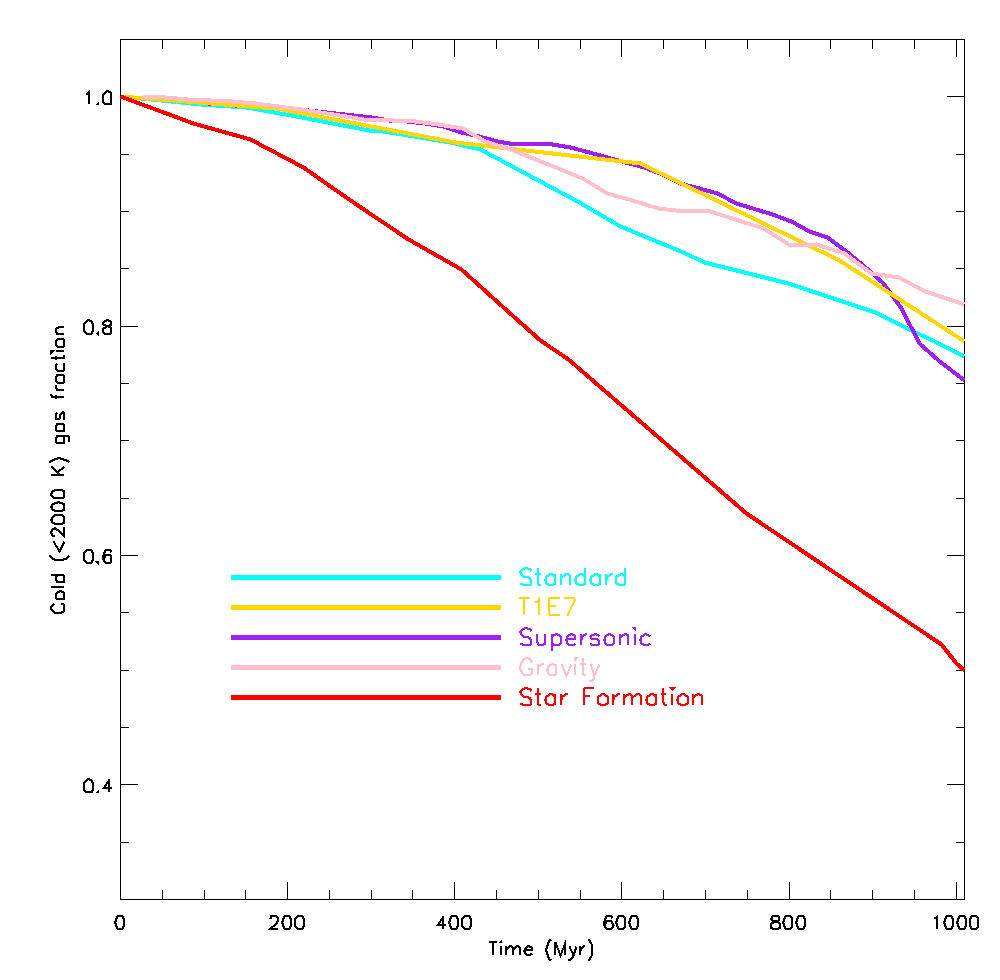}
    \caption{Evolution of the cold gas (i. e. for the gas at $T<2~\times~10^4$ K) fraction as a function of time for all the simulations.
    The solid cyan, yellow, purple, pink and red lines represent the results computed in the Standard, T1E7, Supersonic,
    Gravity and Standard Star Formation model, respectively. }
    \label{fig_cold}
\end{figure*} 

\subsection{The effects of star formation and feedback}
\subsubsection{Star formation history}
\label{sec_sf}
As described in Sect.~\ref{sec_model}, in simulations which include star formation,
the star formation timescale $t_*$ represents an unknown quantity. 
To assess the role of this quantity in shaping the properties of SECCO~1,
we have run a set of star formation simulations with spatial resolution of $\sim$150 pc
(corresponding to maximum refinement level = 8), taking also into account stellar feedback,
in which we considered three initial values for the SF timescale: 0.4 Gyr, 1 Gyr and 40 Gyr.  

The constraints we considered for our models were the observed
stellar mass, the gas mass and the star formation rate (SFR) value, compared to the values of the models at 1 Gyr,
regarded as a reference value for SECCO~1's age, i. e. the time spent in-flight once departed from its parent
galaxy, clearly with the knowledge that this value is uncertain. 

At this time, it was impossible to reproduce at the same time these three constraints.
We came to this conclusion by noticing that with the largest SF timescale value
(i.e. with the lowest SF efficiency), 
although the stellar mass and gas mass values were acceptable, 
the SFR at 1 Gyr was always lower than the observed one by one order of magnitude.
On the other hand, the lowest SF timescale produces too a large stellar mass (of the order of $10^6~M_{\odot}$)
in a short time. 
In the light of these results, we chose for the SF timescale the intermediate value of 1 Gyr. 
Once again, we stress than several unknown quantities may play an important role in our study, including the original 
gas mass and the current age of the system. 

The peculiar nature of SECCO~1 is oulined by its unusual combination of isolation and recent star formation activity. 
We believe that the setup discussed in this section defines a realistic framework,
surely more complete as for its physical ingredients, than the simulations of Sect.~\ref{sec_nosf}, to describe the motion of a  
 star-forming, massive gas cloud in the hot ICM.
It is important to bear in mind that the aim of the present simulations
is to explore the evolution, and not the formation, of SECCO~1, 
and that to investigate how it was originated requires a much different setup, e. g. describing the motion
of an entire galaxy through the ICM \citep{kap09}.

The value assumed for $t_*$ is to be regarded as a representative one,
likely reliable as order-of-magnitude estimate. 
Figure ~\ref{fig_sfr} shows the star formation history of the cloud, computed with the
prescriptions about SF and stellar feedback as described in Sect.~\ref{sec_sf}. 
The two different curves are for simulations computed at different maximum spatial resolution values (76 pc and 152 pc),
and are helpful to appreciate the convergence of our numerical test. 
Numerical convergence seems satisfactory, considering the notorious difficulty
of achieving numerical convergence in AMR simulations (e. g., \citealt{cal15} and references therein).\\
In our simuations, SF begins immediately and, through the whole history, it presents maximum SFR values 
at the earliest times, showing a monotonical decrease and a value at 1 Gyr which is very similar to
the observed estimate. The total stellar mass at 1 Gyr is of $10^6~M_{\odot}$ and the cold gas mass is  
$4~\times~10^6~M_{\odot}$.

\subsubsection{Evolution of the gas and stellar component}
\begin{figure*}
\begin{multicols}{2}
\centering
    \includegraphics[width=9.5cm,height=15cm]{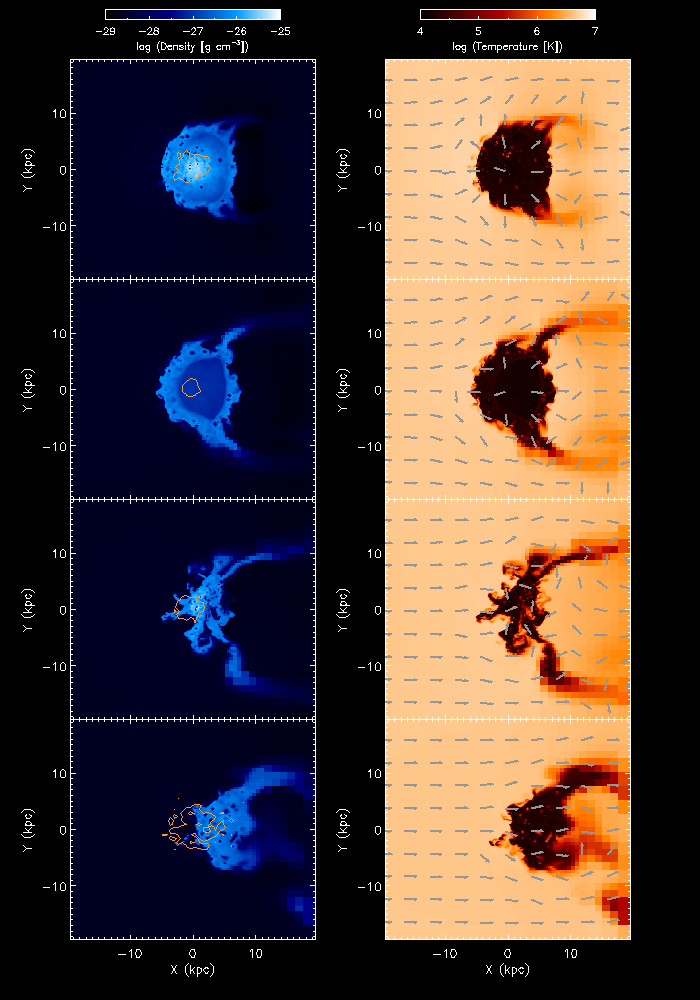}\par
    \hspace{-155pt}
    \vspace{-10pt}
    \includegraphics[width=3.5cm,height=3.99cm]{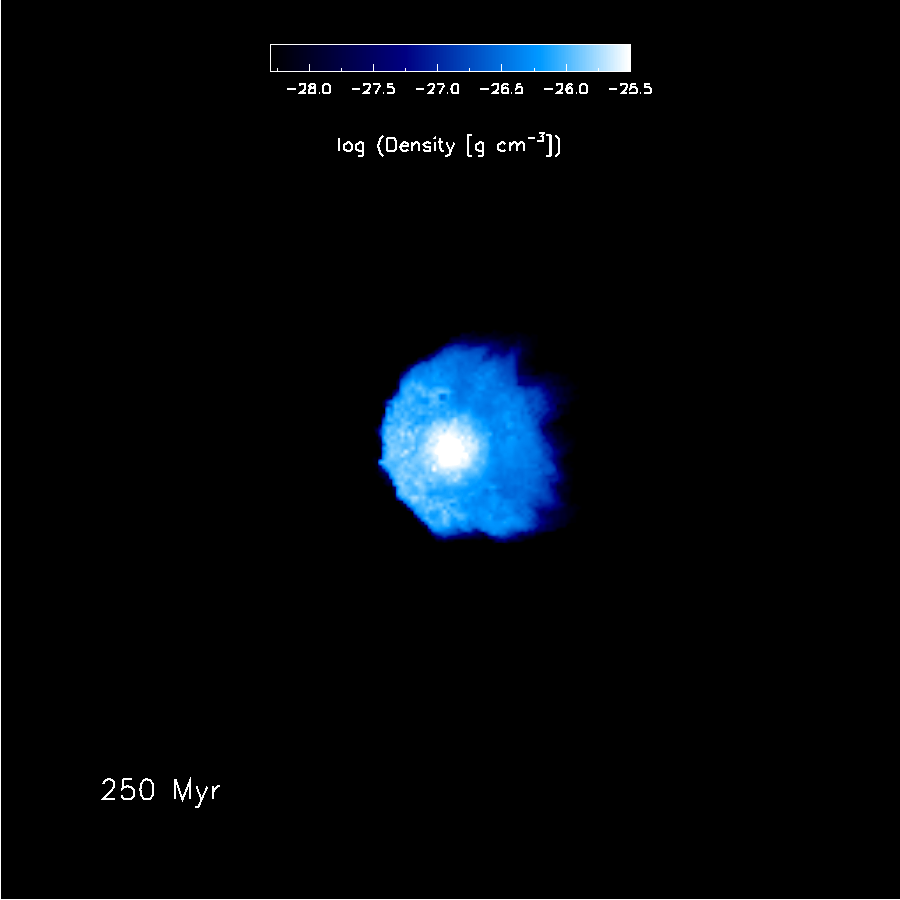}\par
    \hspace{-155pt}
    \vspace{-10pt}
    \includegraphics[width=3.5cm,height=3.99cm]{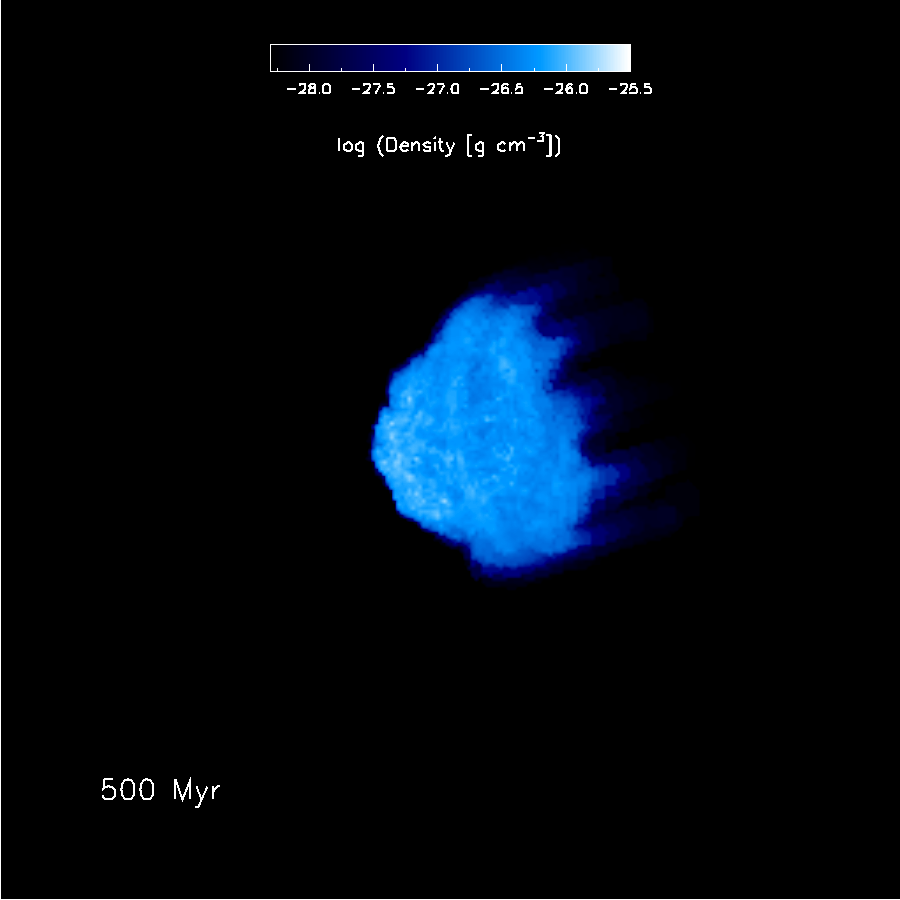}\par
    \hspace{-155pt}
    \vspace{-10pt}
    \includegraphics[width=3.5cm,height=3.99cm]{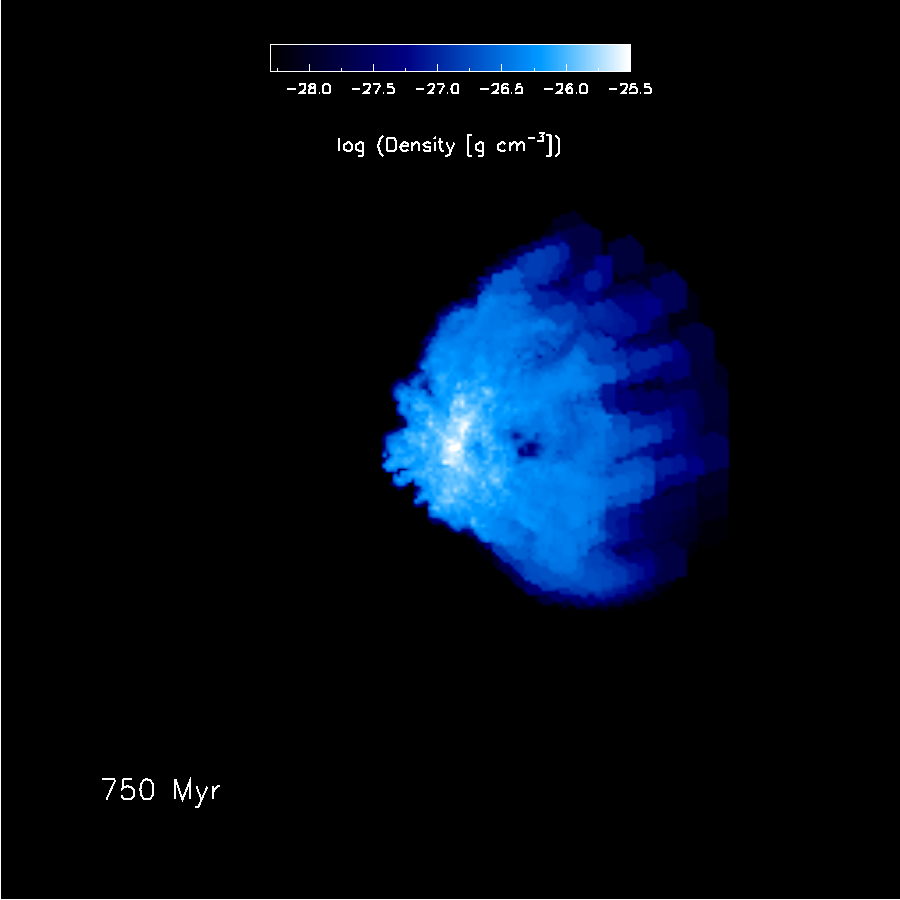}\par
    \hspace{-155pt}
    \vspace{-5pt}
    \includegraphics[width=3.5cm,height=3.99cm]{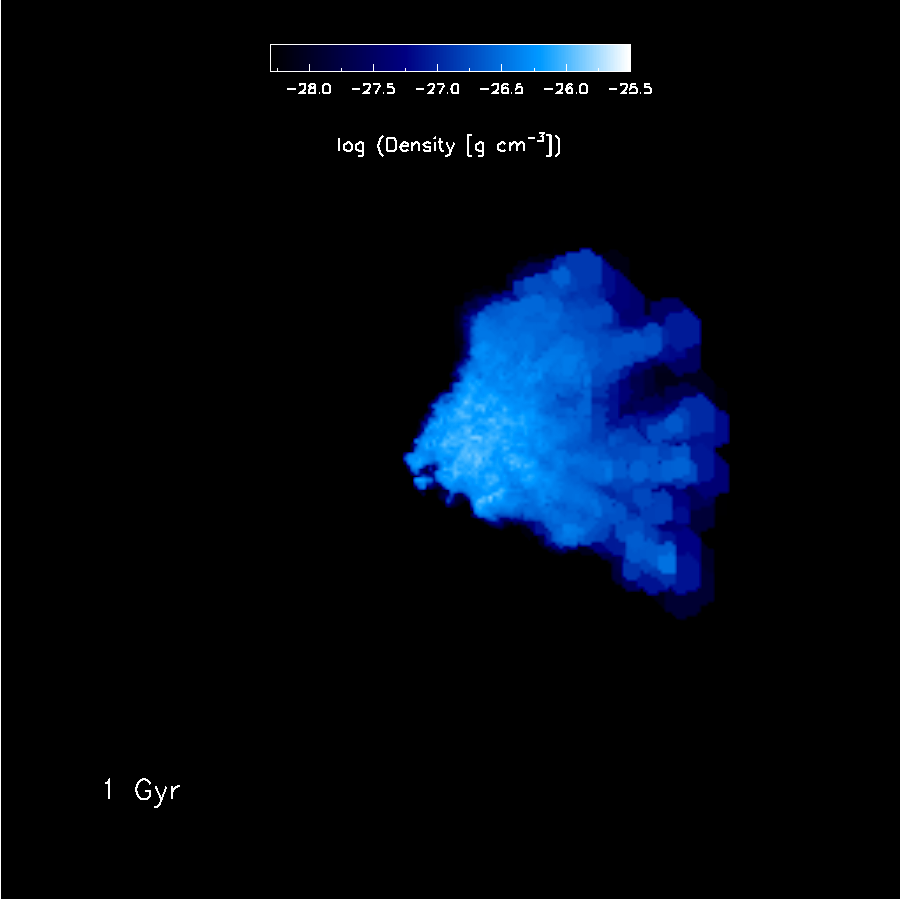}\par
    \end{multicols}
\caption{Two-dimensional gas density (first column), temperature (second column) maps in the x-y plane and three-dimensional
rendered images of the gas density in our simulation with star formation at various time steps. 
In each column, the first, second, third and fourth row describe the model at 250 Myr, 500 Myr, 750 Myr and 1 Gyr, respectively.
The orange contours in the 2D density maps describe the distribution of the stars, and represent
regions which enclose $>50 \%$ of stellar mass. The arrows in the 2D temperature maps are plotted to describe the velocity field of the
gas. }
\label{fig_sf}
\end{figure*}

\begin{figure*}
\includegraphics[width=14.5cm]{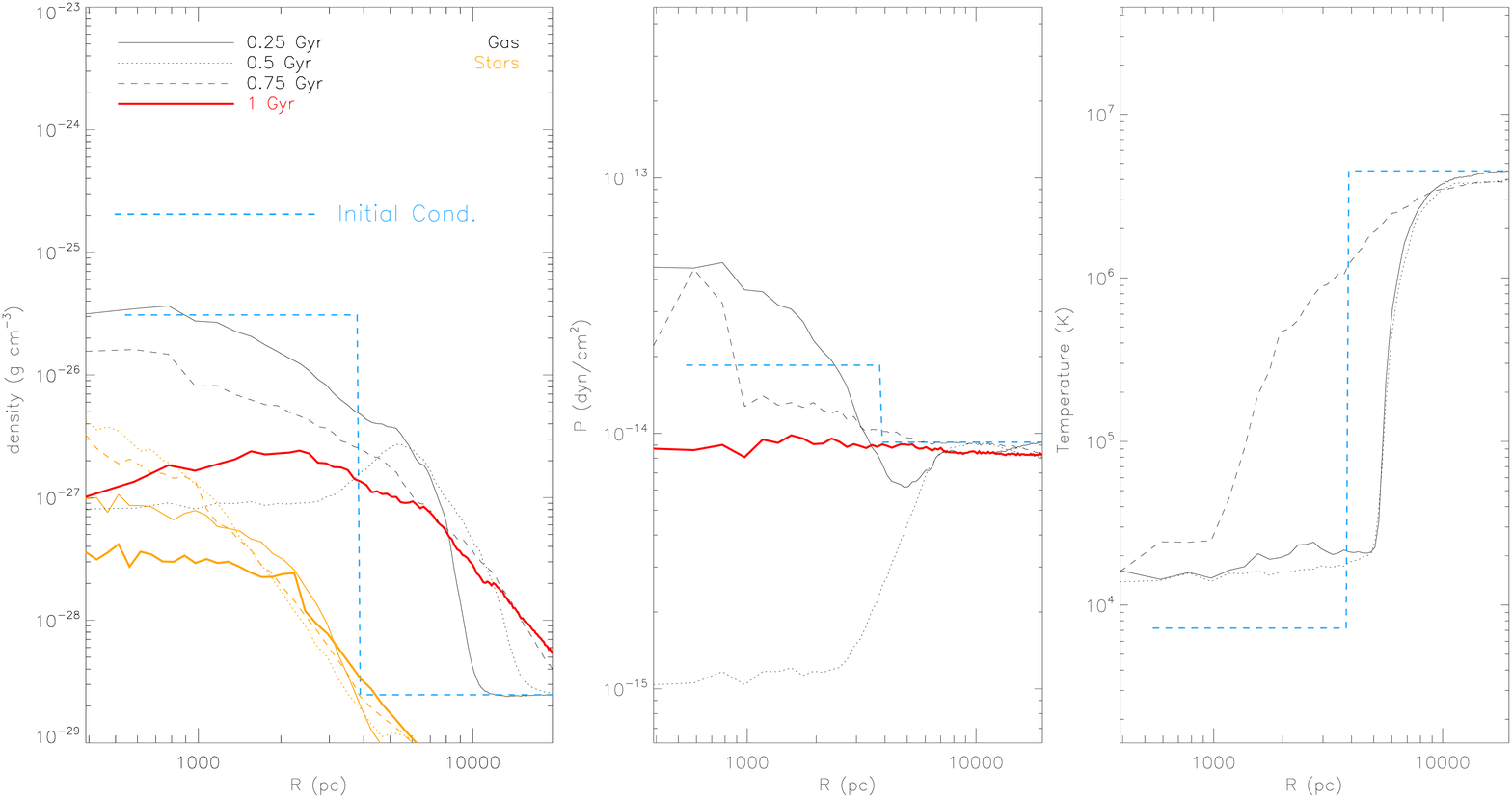}
\caption{Density (left panel), Pressure (middle panel) and temperature (right panel) profiles computed
    for our star formation simulation at different timesteps. 
    In each panel, the blue thick dahed line describes the initial conditions, the thin solid, dotted and dashed lines are the
    profiles computed at 250 Myr, 500 Myr, 750 Myr, respectively, and the solid thick, red line is the final profile, computed at 
    1 Gyr. In the density profile plot, the thin cyan solid, dotted and dashed lines represent the density of the stars
    computed at 250 Myr, 500 Myr, 750 Myr, whereas the thick solid cyan line is for the stellar density profile at 1 Gyr.}
\label{fig_sf_prof}
\end{figure*}

In Figure ~\ref{fig_sf} we show the effects of star formation and stellar feedback
on the evolution of SECCO~1, as traced by 2D density maps, temperature maps and three-dimensional
gas distribution computed at different times.\\
At 250 Myr (top panels) the system still presents its symmetric shape, with the presence
of Kelvin-Helmholtz instabilities already visibile on its sides.
Inside the gas cloud, several hot, rarefied and approximately circular regions are visible which are
due to the effects of stellar feedback. SNe exploding in various regions of
the cloud carve cavities, whose extent depends on the local conditions, in particular on the local
density. SNe exploding in regions far-away from the centre generate larger cavities,
unless the local medium is strongly affected by the compression induced by the growth of the instabilities.
This is in agreement with other grid-based studies of the local effects of stellar feedback in different environments,
i.e. in globular clusters (\citealt{cal15}) and in dwarf speroidal galaxies (\citealt{rom19}). \\
The density inside the SN-driven bubbles is
a few orders of magnitude lower than the one in the centre of the cloud, whereas the temperature
is of the order of $10^6$ K, which is a typical value for the cavity generated by a SN explosion
(e. g., \citealt{cio91}).\\
Although the global profile of the gas cloud is similar to the one of the Standard model (Fig.~\ref{fig_dmap}),
in this case it presents a larger amount of small scale distortions. 
A few SN-driven cavities are visible along both the front and rear profile of the cloud.
In front of such cavities, chimneys of hot gas extending towards the left and right boundaries are visibile in the temperature map.
These features are made of parcels of hot gas moving away from the cloud and, once again,  
are the product of SN explosions, which clearly represent an additional
mechanism though which the initially cold gas is heated and can leave the cloud.
A confined stellar component is already visible, completely embedded in the cold gas. 
The velocity field, plotted in the temperature map, presents a pattern which at this time is not strikingly
different than the one one of the Standard model (Fig. \ref{fig_tmap}). A more perturbed front profile
of the cloud is visible also in the 3D map, which shows an irregular pattern of scattered overdensities,
which are due to the local accumulation of gas generated by the expansion of the SN-driven bubbles and
the multiple encounters of their dense shells. 
At this epoch, the density, pressure and temperature profiles (Fig. ~\ref{fig_sf_prof}) of the model with SF+feedback
are very similar to the ones of the Standard model, confirming that, on the largest scales, the effects of SF are negligible.\\
At t=500 Myr, significant differences with respect to the Standard case are already visible in the 2D density and temperature
maps of Fig.~\ref{fig_sf} , in particular in the fact that they both appear asymmetric. This is visible for instance from
the extent of the two instabilities on the front of the cloud and at its centre,
with the lower one 
(i.e. the one at negative Y values) which is more structured, developed and curled than the upper one, i.e. at positive X-values.
Other signs of asymmetry are in the different extent of the instabilities at the sides of the cloud as
well as along the rear profile. 
These asymetries are due to inhomogeneous SN explosions, occurring in sparse locations within the cloud.
Multiple SN explosions have sensible effects also on the shape of the large cavity at the centre of the cloud.
At this time,  such cavity is globally bell-shaped, but it is
considerably regular in the Standard and 'gravity' cases (Fig.~\ref{fig_dmap} and ~\ref{fig_tmap}),
with limited extension along the x-axis, whereas  
in the SF+feedback simulation it appears distorted and slightly more extended along the x-axis.
Moreover, the density contrast between the inner cavity and its edges is larger in the SF+feedback case than in the Standard.
SN explosions are visible only along the outer shell, as the gas flow in the inner parts is globally divergent
(see the velocity field plotted in Fig. ~\ref{fig_tmap}), hence the conditions for SF (and subsequent SN explosion) are not satisfied.
The 3D map shows that the rear of the gas cloud presents more developed filaments than in the Standard case.
Once again, each SN explosion occurring in various position along the edge of the cloud pushes the gas in different directions,
including outwards of the cloud, thus enhancing the growth of the instabilities, such as the filaments in the back.
At this time, the stellar component is more compact than at the previous time, as visible from the density profile of
Fig.~\ref{fig_sf_prof}, steeper than the one computed at 250 Myr.\\
After 750 Myr, the cloud has undergone considerable evolution, in particular 
its x-y cross-section has considerably reduced. After having expanded, the central regions of the cloud have now re-contracted.  
At this time its asymmetric shape is particularly marked, with
a visibly different aspect of the two longest filaments in its back.
The combined effects of the gravity of the stars and of the compression of the cloud in its centre as due to stellar feedback
have a strong influence on the properties of the cloud. 
Its central density
presents a value considerably higher (nearly by a factor 10) than the one of the Standard case (see 
Fig.~\ref{fig_sf_prof}), whereas its strong central compression is outlined by the pressure profile.
At this time, the central gas density is still of the same order of magnitude of the average density of the real system. 
In the x-y plane, a few hot cavities generated by SNe are still visible, which have exploded in a very narrow region of the cloud, and with small
sizes due to the average high density.
The spatial extent of the stellar component is not much different from the one seen at the previous time. 
However, among the times considered for our plots of Fig.~\ref{fig_sf_prof}, this is the one at which the average stellar density is the largest,
a factor of $\sim 8$ lower than the one of the gas, and remarkably comparable to the central gas density in the Standard case at the same epoch. 
After 1 Gyr, the gas cloud has undergone significant disruption. This is visible from its scattered aspect in the density and temperature maps,
and from its density profile, which in the centre is considerably lower than at the previous time, and also an order of magnitude lower
than in the Standard case. This proves that although self-regulated star formation have caused a contraction of the cloud at previous times,
on a timescale of 1 Gyr the gas cloud is dispersing under its effects. The pressure profile shows that also in this case,
large-scale pressure equilibrium has been reached at this time. 
At this time the scattered, partially disrupted aspect of the cloud is probably more similar to the one of the real system.
In our simulation, star formation is still present and SNe are still exploding, but the stellar component has undergone considerable
expansion with respect to the previous time, with a factor 10 decrease in the central density value.
The scattered stellar component visible in the density map underlines its disgregation, as a result of the disruption of the gas cloud.\\
The presence of star formation and stellar feedback does not seem to affect large-scale properties of the gas cloud, such as
its half-mass radius and velocity linewidth, with values similar to the ones calculated in the Standard simulation without star  formation
(Table~\ref{tab_models}).  On the other hand, 
in the Standard simulation with SF, the evolution of the cold gas fraction is considerably different than the one presented
by all the other cases, with a much stronger decrease but with $50\%$ still present inside the computational volume after 1 Gyr (Fig.~\ref{fig_cold}). 
A fraction of the order of $10\%$ of the gas initially present inside the box has leaked out. Such fraction is increasing with time 
and at 2 Gyr, 80 $\%$ of the initial amount of gas has abandoned the simulated box. 
This prevents us to make quantitative predictions on the physical properties of SECCO~1 at times larger than 1 Gyr, however, our 
study indicates that at this time SECCO~1 is expanding, hence its likely final fate is to be completely disrupted
and disperse in several pieces. 

\subsubsection{Supersonic simulation with star formation}
\label{sec_ssf}
\begin{figure*}
\begin{multicols}{2}
\centering
    \includegraphics[width=9.5cm,height=15cm]{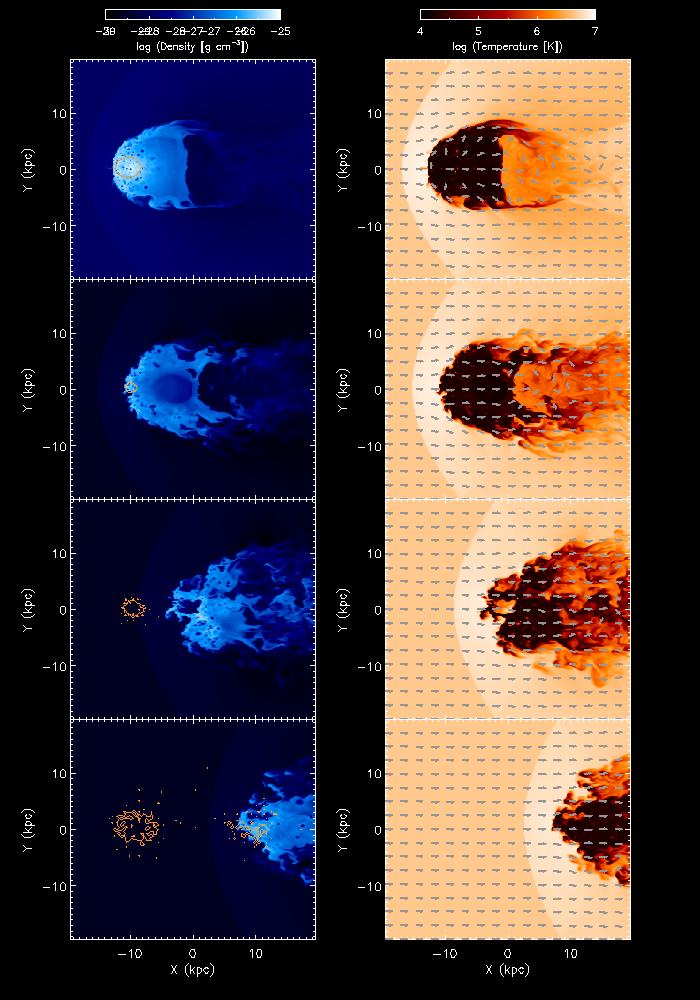}\par
    \hspace{-155pt}
    \vspace{-10pt}
    \includegraphics[width=3.5cm,height=3.99cm]{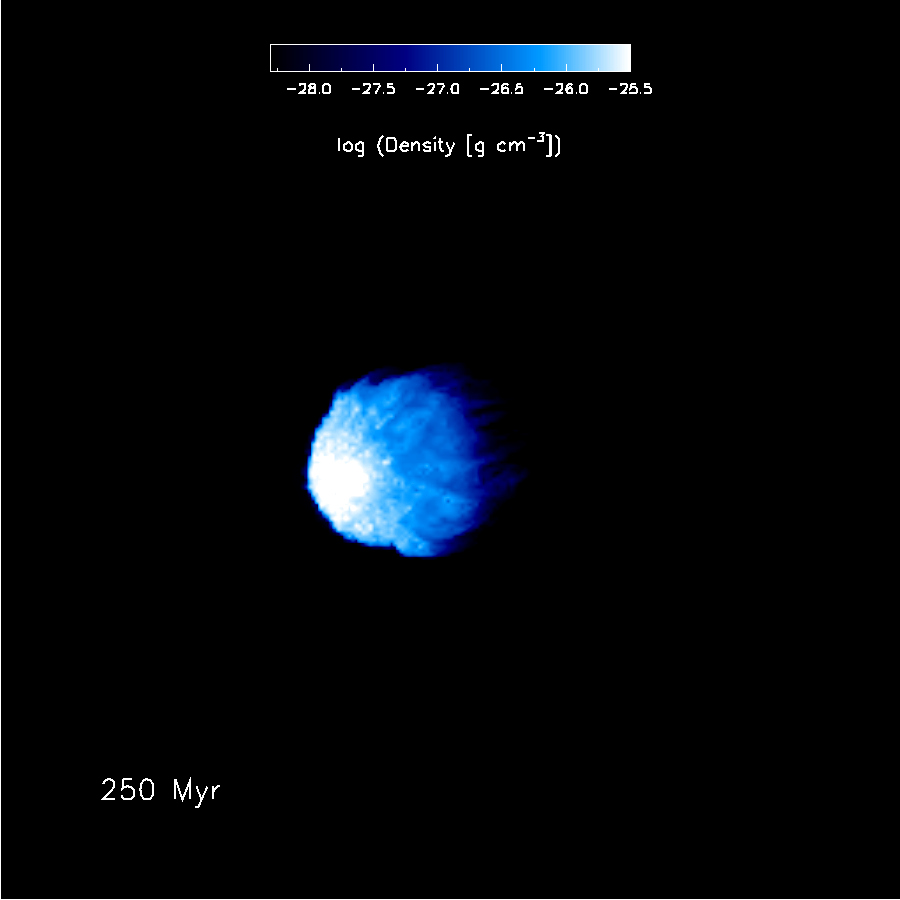}\par
    \hspace{-155pt}
    \vspace{-10pt}
    \includegraphics[width=3.5cm,height=3.99cm]{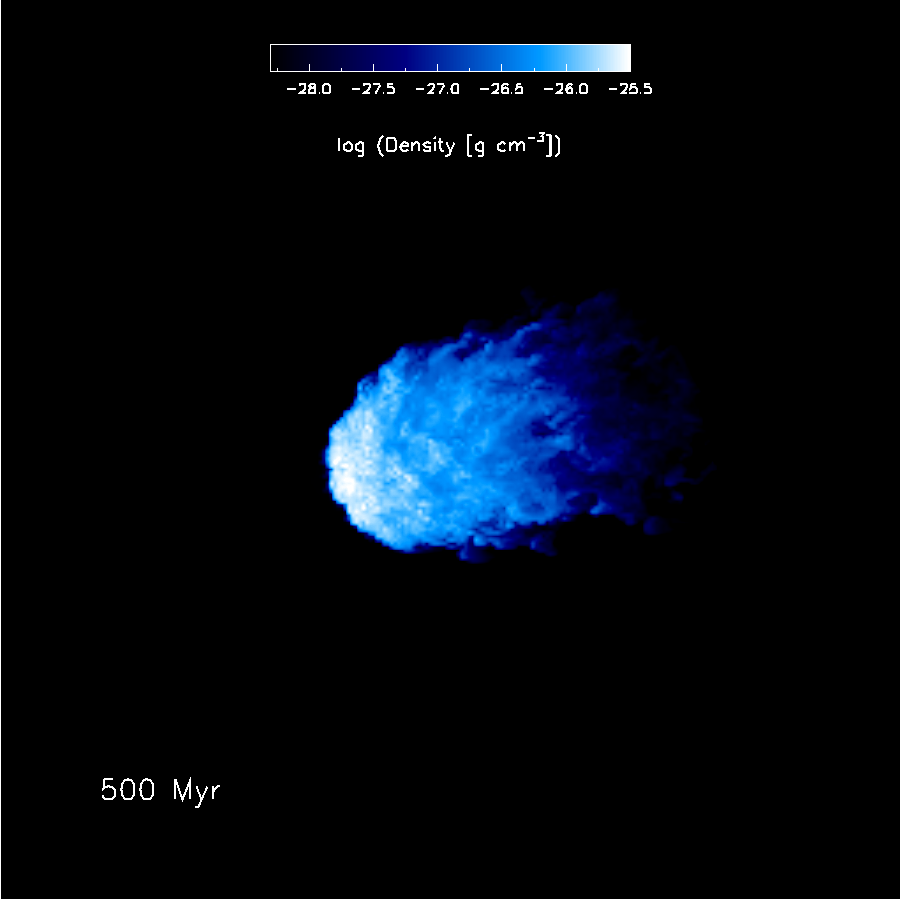}\par
    \hspace{-155pt}
    \vspace{-10pt}
    \includegraphics[width=3.5cm,height=3.99cm]{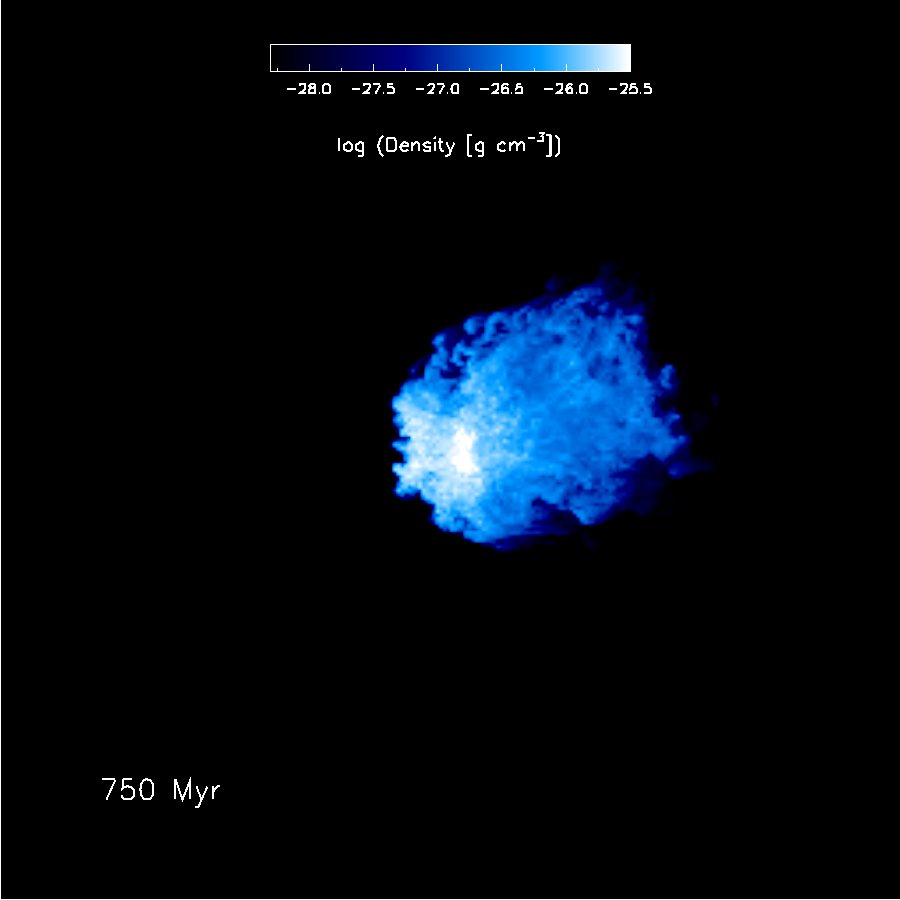}\par
    \hspace{-155pt}
    \vspace{-5pt}
    \includegraphics[width=3.5cm,height=3.99cm]{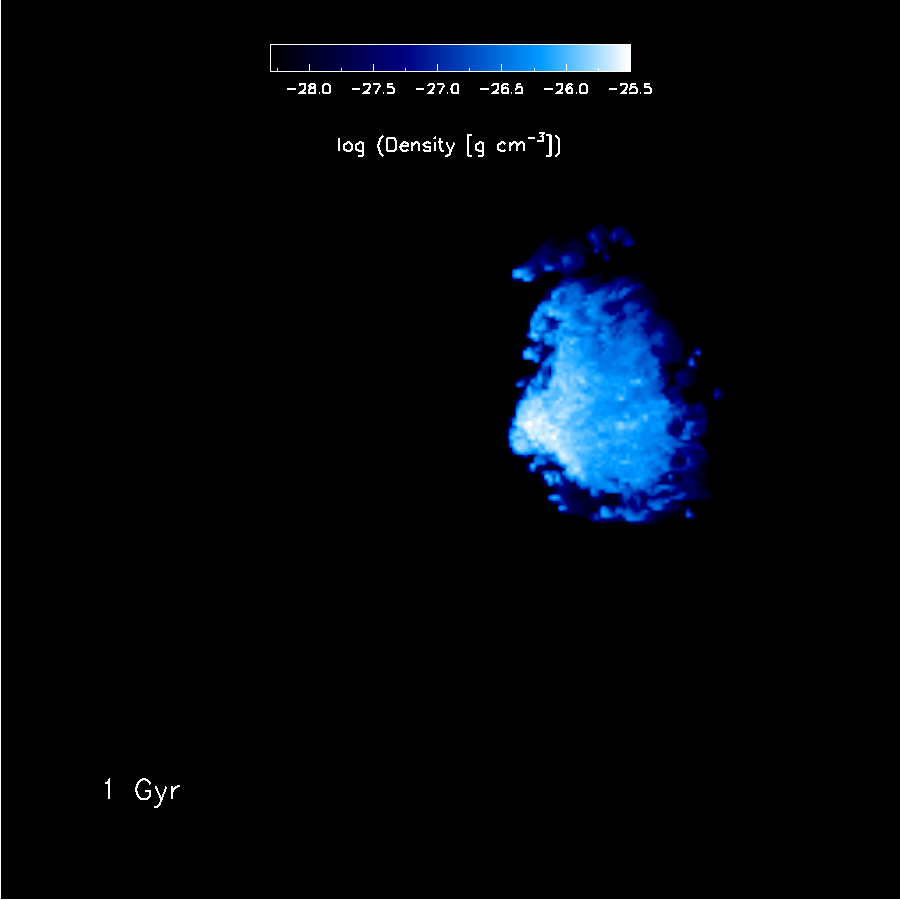}\par
    \end{multicols}
\caption{Two-dimensional gas density (first column), temperature (second column) maps in the x-y plane and three-dimensional
rendered images of the gas density in our Supersonic simulation with star formation at various time steps (see Sect.~\ref{sec_ssf}). 
In each column, the first, second, third and fourth row describe the model at 250 Myr, 500 Myr, 750 Myr and 1 Gyr, respectively.
The orange contours in the 2D density maps describe the distribution of the stars, and represent
regions which enclose $>50 \%$ of stellar mass. The arrows in the 2D temperature maps are plotted to describe the velocity field of the
gas.}
\label{fig_ssf}
\end{figure*}

In Sect.~\ref{sec_nosf} we have tested the effects of a supersonic
motion on the hydrodynamic evolution of the cloud in a simulation withour star formation.
In this section, we investiigate the effects of a supersonic velocity in a more realistic case, i. e.
in presence of star formation and stellar feedback.
The results are shown in Fig. \ref{fig_ssf}, where we present 2D density and temperature maps and 3D rendered images of the density profile
computed at various times. Also in this case, initially the cloud has been placed closer to 
the left boundary than in the Standard simulation, 
in order to reduce the early leakage of gas from the rightmost boundary. \\
The density maps computed at 250 Myr show the bow shock already seen in
Fig. 1 and 2, which is a typical signature of a supersonic motion.
The back of the cloud is stretched along the x- axis and the extent of the rear filaments is visibly larger than in the Standard model. 
Also in this case star formation sets in immediately and 
the supersonic motion causes a high contraction of the cloud, which produces 
a compact stellar component, highlighted by the orange contour in the top-left panel of Fig.~\ref{fig_ssf}. 
Hot cavities are visible scattered around the cloud as the result of multiple SN explosions, with a visible concentration
at the centre of the cloud, where the gas is particularly dense. \\
At 500 Myr the stars occupy a small region on the head of the cloud, whereas the filamentary tail 
is stretched out to the boundary. 
At 750 Myr the gas is decoupled from the stars, with a distance between the compact stellar component and the most prominent
filaments lying on the head of the cloud of a few kpc.
Such separation grows at later times, to reach a value of $\sim~20$ kpc at 1 Gyr.
At this epoch, some stars are still forming at the head of the cloud.
The remainder of the cloud lies close to the right boundary, 
and a significant fraction of its initial mass is leaking out of the box; this is the reason
why this simulation was terminated at 1 Gyr. 
Overall, the supersonic, star forming simulation is characterised by higher star formation rate
with respect to the Standard model. 
Such enhancement in the star formation rate with is visible in Fig. \ref{fig_sfr}, where the star formation
history of the Standard and Supersonic simulations are compared.

As in the Standard simulation, the presence of stellar feedback does not affect much the line width and size
of the cloud (see Tab.~\ref{tab_models}), whereas the evolution of the cold gas fraction is very similar to
the one computed in the Standard model discussed in Sect.~\ref{sec_sf}.

\subsection{Comparing the simulated stellar system with observations}

Are the observed properties of SECCO~1 compatible with those of the
ensemble of stars of our simulation with SF at the end of the
simulation? To try to answer this question we took the t=1~Gyr
snapshot of the stars in the SF simulation and we attached to each
particle a value of the luminosity appropriate for its age,
interpolating on the theoretical model of a Simple Stella Population
(SSP) with Z=0.008, from the BASTI dataset (\citealt{basti1})\footnote{The quantity that is actually interpolated  from
the model is the V-band mass to light ratio in solar units,
$M/L_V$. Multiplying  the inverse of this number for the mass  of the
particle we obtain the V-band luminosity of each individual particle,
according to its age.}. 

Then we computed several quantities that may be useful for the 
comparison with the observed system, like, for example, the velocity
dispersion, the half-mass radius (computed, for each particle age slice, with
respect to the median values of the three spatial coordinates), the
surface brightness within the half-mass radius, etc., for the entire
synthetic sample and for remarkable age slices. For instance, the most
appropriate subsample to compare with SECCO~1 is the age slice
0.0~Myr$< Age\le$50~Myr, since all the  observed stars of the real
system lie in this age range (\citealt{sand17}; B18).

  \begin{figure*}
   \centering
   \includegraphics[width=\columnwidth]{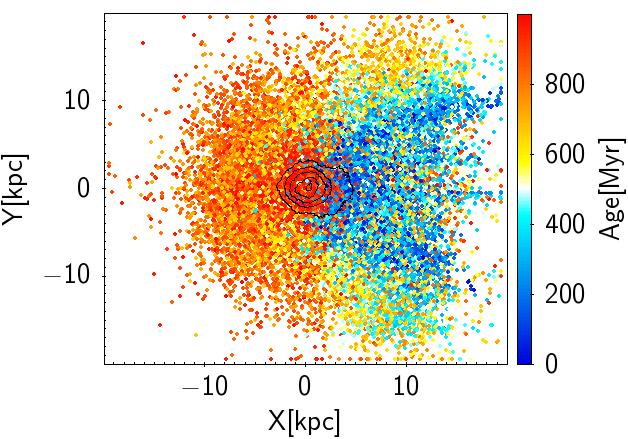}
    \includegraphics[width=\columnwidth]{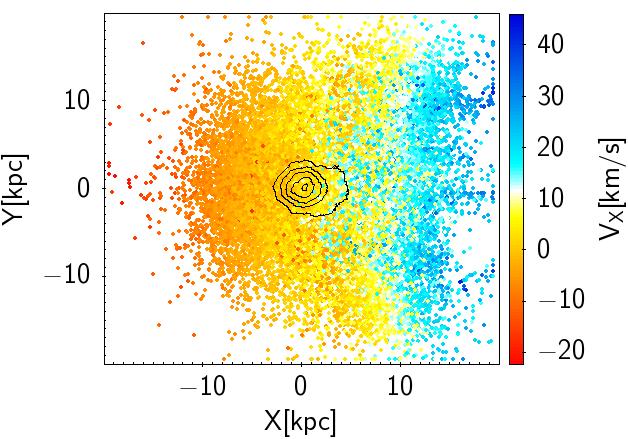}
     \caption{Maps of the final distribution of stars for the simulation with star formation in the X,Y plane. The map in the left panel is color coded according to the age of the stars, the map on the right panel is color coded according to the velocity in the X direction. The superposed contours are density contours in arbitrary units, with a factor of $\sim 1.4$ in density between two consecutive contours, and are intended to show the position and the morphology of the density peak of the distribution.}
        \label{mapstars}
    \end{figure*}
    
Another useful quantity we computed is the virial ratio $\alpha_{vir}$, that, for a spherically symmetric stellar system\footnote{The formula for $\alpha_{vir}$ we used in B18 is appropriate for gas clouds.} with isotropic velocity distribution we define as
 
 \begin{equation}
 \label{vir}
 \alpha_{vir} = 2.32\times 10^5 \left( \frac{3 \sigma^2 R_{hm}}{M}\right)
 \end{equation}
 
 where $ \sigma$ is the velocity dispersion in one direction (in real life along the line-of-sight) in km~s$^{-1}$, $R_{hm}$ is the 3D half-mass radius in kpc,
 and M is the mass in solar masses. 
 According to the Virial Theorem an idealised system like that envisaged above, if gravitationally bound and in virial equilibrium should have $\alpha_{vir} \simeq 1.0$. In practice, given all the involved assumptions and the uncertainties, as a general criteria we can consider a system as gravitationally bound if $\alpha_{vir}\la 1.0$ and unbound if $\alpha_{vir}\gg 1.0$. As a sanity check we used the new dataset for Galactic Globular Clusters (GCs) by H. Baumgardt and collaborators\footnote{\tt people.smp.uq.edu.au/HolgerBaumgardt/globular/parameter.html} (see,  \citealt{baum}, \citealt{hilker}, and references therein) to compute $\alpha_{vir}$ for 152 clusters for which both the 3D half mass radius and the 2D half-light radius ($R_h$; the observable that is typically available in most cases) are provided. The median $\alpha_{vir}$ of this sample is 0.63, with semi-interquartile intervals as small as $\pm 0.05$. There is only one cluster with  $\alpha_{vir}>1.0$, 
 i.e. the remote and anomalous cluster NGC~2419 (\citealt{iba11}, \citealt{muc12}), having $\alpha_{vir}=1.67$. The observed mean $R_{hm}/R_{h}$ for the Galactic GCs is $\simeq 1.5$. Adopting this calibration we can obtain a version of Eq.~\ref{vir} as a function of $R_{h}$, instead of $R_{hm}$ just by replacing the constant 3.0 with $3.0\times 1.5=4.5$, assuming that mass approximately follows light in systems where the contribution of dark matter to the potential should be negligible or null, like GCs or the star forming clouds considered here.
 
 The entire set of simulated stars in the considered snapshot  has $\alpha_{vir}=103$, hence it is clearly unbound, albeit the contribution of the cold gas to the potential may help the system to hold their stars  stronger than what measured by the stellar $\alpha_{vir}$. In this context it is important to note that $\alpha_{vir}$ is driven to values as low as $\simeq 100$ by the oldest stars, that at the end of the simulation have left most of the gas distribution to lag behind (see Fig.~\ref{fig_sf}, lower right panel). The stars with 900~Myr$< Age\le$~1~Gyr have 
$\alpha_{vir}=69.3$ while those with 0~Myr$< Age\le$~50~Myr have  $\alpha_{vir}=2712$.

In Fig.~\ref{mapstars} we show the distribution of the stellar
particles at the end of the simulation in the x,y plane that we take
as the plane of the sky, for convenience, while the z direction,
consequently, is considered as approximately aligned with the
hypothetic line of sight (in computing $\alpha_{vir}$ we adopt
$\sigma=\sigma_z$). The particles are color coded according to their
age (left panel) and to their velocity in the direction of the mean
motion of the cloud (x; right panel). A few density contours
illustrate the position and the morphology of the density peak of the
overall structure. It is interesting to note that a significant 
fraction of the oldest stars maintain a relatively compact and
symmetric configuration, reminiscent of the spherical symmetry of the
original cloud, while the others are diffused over a nearly spherical halo
out to more than 10~kpc from the center of the system. Stars younger
than $\simeq 600$~Myr keep memory of the jellyfish shape of 
the cloud at the epoch of their birth, and finally the youngest stars
traces the dense gas filaments of the legs of the jellyfish at the end
of the simulation.
The mean motion in
the directions perpendicular to the systemic motion (Y and Z) is an
overall expansion, following the axial expansion of the umbrella and
the legs of the jellyfish. When integrated, for instance along the Z
direction, this motion appears as non ordered, since in any given
direction one sees approximately the same number of stars approaching
and receding from us, once the systemic motion is subtracted.  Note
that the velocity dispersions ranges from $\simeq 3$~km~s$^{-1}$ to
$\simeq 8$~km~s$^{-1}$ depending on the age slice and the direction
considered (the gradient shown in Fig,~\ref{mapstars} makes the
dispersion in the X direction larger than that in Y and Z),
corresponding to  3-8 kpc~Gyr$^{-1}$.

Now, focusing on the comparison with SECCO~1 (main properties from Tab.~1 by B18), based on Fig.~\ref{mapstars} and the quantities we computed for the various age slices, in particular the 0~Myr$< Age\le$~50~Myr (``youngest'' hereafter) 
and the 900~Myr$< Age\le$~1000~Myr (``oldest'' hereafter) slices, we can summarise our main conclusions as follows:

\begin{itemize}

\item{} The velocity dispersion ($\sigma_z=4.4$~km~s$^{-1}$) and the half-mass radius ($r_{hm}=4.3$~kpc) of the youngest population are similar to those of SECCO~1. The amplitude of the velocity gradients along the direction of motion is compatible with that of the observed gradients (\citealt{pap_muse}).  In the simulation, the youngest stars form along the legs of the gaseous jellyfish, in filamentary structures similar to SECCO~1 as well as the other candidate SECCO~1-like systems identified by \cite{sand17}.

\item{} The overall distribution  of the simulated stars, on the other hand, seems quite different. Even when only the stars younger than 10 Myr are considered,  they are distributed over a very large area, of the order of 20~kpc$\times$20~kpc, while the distance between the two pieces of SECCO~1 
(main body and  and secondary body, see B18) is only 7.3~kpc, and B18 were able to find only one additional  candidate H$_{\rm II}$ region possibly associated to SECCO~1 over a surveyed circular area of radius=~20~kpc around the system.

\item{} The total luminosity of the youngest population is $4.4\times 10^5~{\rm L_{V,\sun}}$, that is $\simeq 2.6$ times smaller than the total luminosity of SECCO~1. Even rescaling the luminosity of the youngest population by this factor the average surface brightness within the 2D half-mass radius is $\mu_{V,e}=30.5$~ mag/arcsec$^2$, a factor of $\sim 4-12$ lower than what observed in SECCO~1 secondary body ($\mu_{V,e}=28.9$~ mag/arcsec$^2$) and main body ($\mu_{V,e}=27.8$~ mag/arcsec$^2$)

\item{} The SFR in the last 50~Myr of the simulation is $0.6\times 10^{-3}$ $M_{\odot}$ yr$^{-1}$ in excellent agreement with the current SFR measured in SECCO~1.

\item{} As said, the oldest stars are able to preserve a spheroidal morphology, and display the smallest size ($r_{hm}=2.1$~kpc), the lowest velocity dispersion ($\sigma_z=4.4$~km~s$^{-1}$), and the brightest average surface brightness ($\mu_{V,e}=29.7$~ mag/arcsec$^2$) of any other age slice of the final snapshot  of the simulation. On the other hand, for a combination of the spatial diffusion and of the behaviour of $M/L_V$ with age, the intermediate-age slices are those with the faintest surface brightness (e.g.,$\mu_{V,e}=33.5$~ mag/arcsec$^2$ for the slice 450~Myr$<Age\le$550~Myr), hence they would be the hardest to detect observationally.

\end{itemize} 

In the context emerging above, it is intriguing to note that there is a low-surface brightness (SB) dwarf spheroidal (dSph) in the vicinity of SECCO~1,
at a projected distance compatible with that between the density peak of the oldest stars and the youngest stars of the simulation (\citealt{sand17}; B18). 
Much deeper observations than those currently available are required to verify if SECCO~1 and the nearby dSph are immersed into a wedge-shaped low-SB distrubution
of intermediate age stars as the one we see in the end state of our simulation.

\subsection{Discussion and comparison with other results}

So far, several theoretical studies have been dedicated to the evolution of
ultra-diffuse systems with some of the features presented by SECCO~1 
in various environments (e. g., \citealt{plo15}; \citealt{tay18}; \citealt{hau19} and references therein, \citealt{sil19}). 
Hovever, there were not many previous attempts to model the evolution of 
isolated, diffuse, initially unbound objects with no dark matter content
and with size and mass comparable to the ones of SECCO~1.  
In this Section, we discuss a few of such extant
studies and, where possible, compare their results to ours. \\
By means of an analytic approach, \cite{bur16} discussed the possibility that  
starless, dark HI clumps might be supported by external pressure in the surrounding intercluster medium.
From their surface density, they predicted that some of the HI dark clumps detected in the Virgo Cluster by the ALFALFA HI Arecibo survey
might be on the cusp of forming stars. This conclusion was based on the fact that  many systems were lying
above the HI-H2 transition threshold density, which in principle might mark the threshold for star formation,
estimated to be between 4 and 10 $M_{\odot}/$pc$^2$.
The average surface density of SECCO~1 is well below such values (B18). The fact that SECCO~1 is forming new stars
puts into question that a threshold value for star formation may really exist.

Past works have shown that the adoption of a density threshold of 100 $cm^{-3}$ was necessary to produce bulgeless
dwarf galaxies in simulations (e. g., \citealt{gov10,pil11}). Since then, in a great multitude of paper star formation was modelled
through the adoption of a density threshold criterion, sometimes with much lower values (0.1  $cm^{-3}$) and with resolution lower than the maximum resolution used here (e. g., \citealt{pil12,few14}). 
At the scales probed by our simulations (76 pc), most of the times the gas is compressed to density values lower than
0.1 $cm^{-3}$. In simulations where star formation occurs mostly in gravitationally bound regions, such density
value can be reached easily on large scales, e. g. even a few hundred pc (e. g., \citealt{pil11}).
We cannot exclude that at higher resolution larger density values could be reached, however another possibility is
that, in an unbound system such as SECCO 1 where star formation is activated by external compression from an external, 
hot medium, it can occur at lower average density values than in typical star-forming galaxies. 

Observationally, the 
spatial properties of local star forming galaxies are studied by means of the Schmidt-Kennicutt (SK) relation (\citealt{sch59,ken98})
between SFR and gas surface density in entire galaxies or in sub-galactic region. 
The resolved SK law (e.g. \citealt{ken89}; \citealt{mar01}) 
involves the gas and the SFR surface densities measured either in kpc- or sub kpc-scale regions or their radial profiles and shows
a break of correlation in low density environments, such as dwarf galaxies and the outskirts of spirals \citep{ken07} 
which was interpreted as a drop in the star formation efficiency at a surface density threshold density of $\sim 10~M_{\sun}/$pc$^2$. 
By using a sample of local disc galaxies, from the study of the volume gas and SFR  densities, \cite{bac19a}
showed that a single-slope, volumetric SK law is recovered down to a gas density of $\sim~0.04$ $cm^{-3}$ at scales of
a few $\sim 100$ pc. 
This implies that the break observed from the classic, two-dimensional SK law could due to a flaring effect which is common in disc galaxies
\citep{elm15,bac19b}, 
and argues against the existence of a star formation density threshold, at least down to a density value $<0.1$ cm$^{-3}$, i.e.
lower than the ones previously regarded as reference threshold values. 
To shed further light on this fundamental aspect, more studies are needed of the spatial
star formation properties of low-density systems, of which SECCO 1 surely represents a prototype, perhaps of a peculiar subclass. 
Another interesting aspect concerns the presence of molecular gas as a necessary ingredient to ignite star formation
(\citealt{kru12}; \citealt{glo12}; \citealt{bac19a}). 
In a comprehensive study of the Schmidt-Kennicutt SF law in local galaxies, \cite{big08} evidenced
a transition from HI-dominated SF galaxies to molecular-dominated ones at $9~M_{\sun}/$pc$^2$.
SECCO~1's gas reservoir is most likely dominated by HI, and it lies well below this density value. 
The existence of an extreme system such as SECCO~1 
might be the proof that SF can occurr and be sustained even in systems with no (or very little) molecular gas. 
In order to have a final proof and to have constraints on the molecular gas content, very deep infrared observations will be required. 
Regarding theoretical studies of the possible origin of systems like SECCO~1 and their star formation activity, 
\cite{kap09} performed smoothed-particle hydrodynamical simulations with radiative cooling, star formation and stellar feedback,
of the effects of ram-pressure stripping of the motion of a large disc galaxy in a cluster environment.
They have tested various values for the relative velocity of the galaxy and the ICM and for the density of the ICM.
In general, they found that the effects of the ram-pressure is to strip several gas clouds, liying in extended
wakes donwstream of the parent galaxy. 
The gas stripped from the disc can reach distances of several hundreds kpc from the parent galaxy in a few hundred Myr.
Star formation occurs at any time in the stripped gas clouds and some of these clouds, at any distance from the parent galaxy along the wake,
may be still forming stars after 500 Myr, and hence they may host ionized cavities such as the HII regions observed in SECCO~1 and
found in this study. Also in this study, the star-forming knots are generally associated with underlying older stellar populations. 

\section{Conclusions} \label{sec_conclusions}
In this work, we have presented a novel suite of 3D, high-resolution hydrodynamic simulations of the motion of a cold gas cloud 
throughout a background, uniform distribution of hot gas, representing the intra cluster medium (ICM) in Virgo. 
In our simulations, the cold gas cloud was assumed to be initially spherical and uniform. 
Our focus was to study its evolution after the cloud is already far away from its parent galaxy, not anymore subjected to the effects of any external gravitational field.
The results presented in this work expand the study 
presented in B18 in which one simulation was run, with only one reference value for a few key parameters was adopted, i. e.
an external ICM temperature of $5\times 10^6$ K and a cloud speed of 200 $km/s$.
In the present study, we have tested the effects of a hotter ICM (i. e. $T_{ICM}=10^7$ K),
of a larger relative speed ($v=400$ km/s) between SECCO~1 and the ICM on the evolution of the system,
and of the self gravity of the gas, which is expected to stabilise further the cloud against the development of hydrodynamic instabilities. 
To address all of these issues, first we have run 
a few hydro simulation without taking into account star formation and stellar feedback, but including the effects of
a hotter ICM, a larger cloud speed with respect to the 3D model presented in B18 (regarded as the Standard model),
and finally one simulation with the same parameters as in B18 but including the self gravity of the gas. 
After such pure hydrodynamic, radiative runs, we have improved our model and considered  
star formation and stellar feedback to study their effects on the evolution of the gas cloud.\\
Our main results can be summarised as follows.
\begin{itemize}
\item We have deepened our study of our Standard simulation  (with parameters same as in B18),
presenting new 3D rendered images of the gas distribution at various evolutionary times.
Our results have shown that a timescale of $\sim$0.5 Gyr is enough for a system such as
SECCO~1, to evolve from an initial spherical shape (as assumed here) to an umbrella-like or jellyfish-like shape.
Our study of the evolution of density and pressure profiles have outlined an oscillating evolution of the central density value,
and that pressure equilibrium with the external environment was reached already after 0.5 Gyr of evolution. 
\item With respect to the Standard model, a larger temperature value ($10^7$ K) assumed for the ICM
produced, at any time,  a more compressed gas cloud. In the first 0.5 Gyr, the progressive increase of the inner density and pressure is stronger. 
The aspect of the instabilities at the edges of the cloud is different, in that they appear more elongated, spatially separated and curled
with respect to the Standard case.
\item In the case of a supersonic relative velocity between the cloud and the ICM, the cloud undergoes a
strong compression at its front, and its stretching mostly occurs along the direction of
motion, which in our reference frame is along the x-axis, whereas the expansion of its back is much more limited along the two y- and z- axis.
Similar to the assumption of a large ICM temperature, 
the choice of large velocity of the cloud
helps increasing its capability to retain its initial cold gas reservoir on long timescales, of the order of $1 Gyr$. 
\item In our hydrodynamic simulation in which the self-gravity of the gas cloud was taken into account,
on large scales and on a long timescale the properties of the cloud are very similar to
the ones of the Standard model. This experiment was thus important to directly show that
the role of gravity in the long-term evolution of the cloud and at the largest scales is negligible.
Gravity does not prevent the growth of large scale
instabilities, such as the elongated filament along its trail, but it
hinders the development of small-scale instabilities, in particular on the front of the cloud.
\item For each model, the cloud size and line width were computed, expressed by half-mass radius as obtained from
the two-dimensional density profile and as the square root of the sum of squares of the 1D velocity dispersion and sound speed, respectively.
 Since the temperature floor assumed here is characterised by sound speed comparable to the observed line width,
for the latter quantity only upper limits were calculated. 
All models show similar sizes, and supersonic models show upper limits on the line width higher than the subsonic ones. 
\item All the models which do not include star formation show a similar evolution of the cold gas fraction with time,
and similar values after 1 Gyr. Final values range between 0.75 and $\sim$0.8, consistent with the Standard
model. 
\item In our simulation with star formation,
SF begins immediately, it presents its maximum values
at the earliest times and  decreases monotonically with time. 
After 1 Gyr, the SFR value is comparable with  
the observed estimate, but the total stellar mass is of $10^6~M_{\odot}$,
larger than what observed in SECCO~1 .
The end state of the resulting (unbound) stellar system presents both differences as well as intriguing similarities with the observed structure. 
Throughout its entire evolution, due to inhomogeneous SN explosions,
the appearence of the cloud is much more asymmetric than in the previous cases.
The combined effects of the gravity of the stars and of the compression of the cloud in its centre as due to stellar feedback
have a strong influence on the properties of the cloud.
Due to an enhanced compression of the gas in the outskirts, stellar feedback has the global effect of
facilitating the develompment of instabilities. After 1 Gyr, $50 \%$ of the initial cold gas is still present
in the box.\\ Also a supersonic, star forming model was presented. In the case of supersonic motion, star formation is
enhanced with respect to the Standard case. Moreover, the spatial separation between the stellar component and the gas reaches values
of several kpc already after 0.5 Gyr. 
\end{itemize}
In summary, our simulations confirm that in a variety of plausible conditions,
the survavibility of SECCO~1 is granted on timescales of the order of several hundreds Myr. 
Although our simulations do not allow us to quantitatively address the evolution of SECCO~1 on a timescale larger than 1 Gyr (mostly because
most of the gas leaks out of the volume), the most likely final fate of both the stellar and gas comoponents is to  expand further and to be dispersed 
in the Virgo cluster.\\
The James Webb Space Telescope will provide the first opportunity to test if the SF scenario and the
distribution of stars envisaged 
in our simulations are (at least partially) similar to those of the real systems, 
as a significant part of them may lie buried below the sensitivity limits of the
presently available observations.

In the future, it will also be important to investigate futher on the formation of SECCO 1. 
A plausible hypothesis is that it was pressure-stripped from a parent galaxy. 
Such a scenario can be investigated in a model by means, e. g., of a wind tunnel simulation in which a disc galaxy
is exposed to a fast, low-density gas such as the one modeled here (\citealt{kap09,rug17}). 

\section*{Data Availability} 
The data that support the findings of this study are available from the corresponding author, F. C., upon reasonable request.

\section*{Acknowledgements} 
We acknowledge the MoU INAF-CINECA (Accordo-Quadro 2017-2019) 
for the availability of high-performance computing and support.
An anonymous referee is acknowledged for a constructive report and several useful comments.  
We are grateful to C. Bacchini for several interesting discussions. 
FC acknowledges support from grant PRIN MIUR 2017 - 20173ML3WW\_001 and 
from the INAF main-stream (1.05.01.86.31) 










\bsp	
\label{lastpage}
\end{document}